\documentclass{aa}
\usepackage[pdfpagelabels=false]{hyperref}
\hypersetup{colorlinks=true,
            linkcolor=blue,
            citecolor=blue,
            filecolor=blue,
            urlcolor=blue}
\usepackage{graphicx}	
\usepackage{bm}		
\usepackage{pdflscape}	
\usepackage[T1]{fontenc}
\usepackage{ae,aecompl}
\usepackage{newtxtext,newtxmath}
\usepackage{orcidlink}
\usepackage{framed}
\usepackage{comment}
\usepackage{enumerate}
\bibpunct{(}{)}{;}{a}{}{,} 

\DeclareRobustCommand{\r}[1]{{\rm #1}}
\DeclareRobustCommand{\Sec}[1]{Sec.~\ref{#1}}
\DeclareRobustCommand{\Tab}[1]{Table~\ref{#1}}
\DeclareRobustCommand{\Fig}[1]{Fig.~\ref{#1}}
\DeclareRobustCommand{\Eq}[1]{Eq.~(\ref{#1})}

\newcommand{\Tr}{T_{\r{radio}}}

\newcommand{\LXIII}{\mathcal{L}_\r{x,III}}
\newcommand{\Tb}{T_{21}}
\newcommand{\Tbb}{\bar{T}_{21}} 
\newcommand{\xH}{x_{\r{HI}}}
\newcommand{\be}{\begin{equation}}
\newcommand{\ee}{\end{equation}}
\newcommand{\bea}{\begin{eqnarray}}
\newcommand{\eea}{\end{eqnarray}}

\newcommand{\rd}{{\rm d}}
\newcommand{\OmC}{\Omega_{\r{c}}}
\newcommand{\OmB}{\Omega_{\r{b}}}
\newcommand{\OmM}{\Omega_{\r{m}}}

\begin{document} 

\title{The EDGES measurement {\it disfavors} an excess radio background during the cosmic dawn}

\author{
	Junsong Cang \inst{1,2,3,4} \thanks{cangjunsong@outlook.com}
	\orcidlink{0000-0002-0061-0728}
	\and
	Andrei Mesinger \inst{1}
	\orcidlink{0000-0003-3374-1772}
	\and
	Steven G. Murray \inst{1}
    \orcidlink{0000-0003-3059-3823}
	\and
	Daniela Breitman \inst{1}
    \orcidlink{0000-0002-2349-3341}
	\and \\
	Yuxiang Qin\orcidlink{0000-0002-4314-1810} \inst{5}
	\and
    Roberto Trotta \inst{2,6}
    \orcidlink{0000-0002-3415-0707}
}

\institute{
    Scuola Normale Superiore, Piazza dei Cavalieri 7, 56126 Pisa, Italy
    \and
    Theoretical and Scientiﬁc Data Science, Scuola Internazionale Superiore di Studi Avanzati (SISSA), Via Bonomea 265, 34136 Trieste, Italy
    \and
    Key Laboratory of Particle Astrophysics, Institute of High Energy Physics, Chinese Academy of Sciences, Beijing, 100049, China
    \and
    School of Physical Sciences, University of Chinese Academy of Sciences, Beijing, 100049, China
    \and
    Research School of Astronomy and Astrophysics, Australian National University, Canberra, ACT 2611, Australia
    \and 
    Physics Department, Astrophysics Group, Imperial College London, Prince Consort Road, London SW7 2AZ, UK
}
    
\date{Received XXXX; accepted XXXX}

\abstract{

In 2018 the EDGES experiment claimed the first detection of the global cosmic 21cm signal, which featured an absorption trough centered around $z \sim 17$ with a depth of approximately -500mK.
This amplitude is deeper than the standard prediction (in which the radio background is determined by the cosmic microwave background) by a factor of two and potentially hints at the existence of a radio background excess.
While this result was obtained by fitting the data with a phenomenological flattened-Gaussian shape for the cosmological signal,
here we develop a physical model for the inhomogeneous radio background sourced by the first galaxies hosting population III stars.
Star formation in these galaxies is quenched at lower redshifts due to various feedback mechanisms,
so they serve as a natural candidate for the excess radio background hinted by EDGES,
without violating present day measurements by ARCADE2.
We forward-model the EDGES sky temperature data, jointly sampling our physical model for the cosmic signal, a foreground model, and residual calibration errors.
We compare the Bayesian evidences obtained by varying the complexity and prior ranges for the systematics.  We find that the data is best explained by a model with seven log-polynomial foreground terms, and that it requires calibration residuals.  
Interestingly, the presence of a cosmic 21cm signal with a non-standard depth is decisively disfavored.
This is contrary to previous EDGES analysis in the context of extra radio background models, serving as a caution against using a “pseudo-likelihood” built on a model (flattened Gaussian) that is different from the one being used for inference.
We make our simulation code and associated emulator publicly-available.}

\maketitle
\defcitealias{Bowman:2018yin}{B18}
\defcitealias{Sims:2019kro}{SP20}

\section{Introduction}

By mapping out the first half of our observable Universe, the cosmological 21cm signal from neutral hydrogen promises to revolutionize astrophysics and cosmology.
Experimental efforts to detect this signal fall into two categories.
Interferometers like the Square Kilometre Array (SKA; e.g. \citealt{Mellema:2012ht}), Hydrogen Epoch of Reionization Array (HERA; \citealt{HERA:2021noe}), Low Frequency Array (LOFAR; \citealt{LOFAR:2013jil}) and Murchison Widefield Array (MWA; \citealt{Tingay:2012ps})
are dedicated to measuring the spatial fluctuations of the 21~cm brightness temperature $\Tb$.
Alternatively, global experiments including EDGES~\citep{Bowman:2018yin}, 
SARAS~\citep{Singh:2017syr,Bevins:2022ajf},
REACH \citep{deLeraAcedo:2022kiu},
MIST~\citep{Monsalve:2023lvo}
and LEDA~\citep{Greenhill:2012mn} aim to detect the spatially-averaged (global) 21~cm signal $\bar T_{21}$,

The first detection of the cosmic 21~cm signal was claimed by the EDGES experiment in 2018 (\citealt{Bowman:2018yin}, hereafter \citetalias{Bowman:2018yin}),
which measured a flattened $\Tbb$ absorption profile centered around $z\sim 17$ with an amplitude of $500^{+500}_{-200}$mK (at 99\% credible interval).  A subsequent Bayesian reanalysis of the entire signal chain by \citet{Murray:2022uhg} confirmed the presence of such a signal in the data.
This signal is about twice as deep as the maximum amplitude allowed by standard models in which baryons cool at most adiabatically and the radio background is dominated by the cosmic microwave background (CMB).
As has been pointed out in many previous studies, a cosmological origin for the EDGES signal necessitates either extra gas cooling, 
e.g. via interaction between dark matter (DM) and baryons~\citep{barkana2018possible,Munoz:2018pzp,Berlin:2018sjs},
early gas-CMB decoupling in early dark energy models~\citep{hill2018can},
or an excess radio background component in addition to the CMB~\citep{Feng:2018rje, ewall2018modeling, mirocha2019does,Ewall-Wice:2019may, Reis:2020arr}.

It is very challenging to perform accurate inference from the global 21cm signal.  Astrophysical and terrestrial foregrounds are 4 to 6 orders of magnitude brighter than the cosmological signal (e.g. \citealt{Pritchard:2011xb,Hibbard:2023sdv}) and unlike interferometers, global 21cm experiments cannot clean the signal by comparing different pointings and sightlines (e.g. \citealt{Nasirudin:2020qvr}).
In addition to astrophysical foregrounds,
other systematics such as the Earth's ionosphere (e.g. \citealt{Datta:2014xla,Shen:2020jwt}),
antenna beam effects (e.g. \citealt{Mahesh:2021rly,Sims:2022nwg}), and signal-chain defects (e.g. \citealt{Monsalve:2016xbk,Murray:2022uhg}) can further obscure the signal and potentially lead to false reconstructions~\citep{Tauscher:2020wso,Hibbard:2023sdv}.
Therefore, the interpretation of the EDGES results requires detailed characterisation of these contaminants together with the cosmic signal, with all of the resulting parameters being constrained through Bayesian inference.

The 21cm signal reported in \citetalias{Bowman:2018yin} was obtained by fitting the EDGES sky temperature data with a flat-Gaussian $\Tbb$ profile and a 5-term polynomial foreground model.
Subsequent analysis in~\citealt{Hills:2018vyr,Singh:2019gsv} showed that the best-fit profile found in \citetalias{Bowman:2018yin} requires an un-physical foreground,
and that the EDGES data can also be fit with multiple $\Tbb$ shapes that are different from those used in \citetalias{Bowman:2018yin}.
\citealt{Sims:2019kro} (hereafter \citetalias{Sims:2019kro}) extended these analyses by considering a set of models with various combinations of the cosmic signal, residual calibration systematics and foregrounds.
After comparing the Bayesian evidence for a total of 128 models,
\citetalias{Sims:2019kro} 
found that models that are not strongly disfavored included a flattened Gaussian cosmic signal with a non-standard depth and sinusoidal calibration residuals.
Subsequent data from the SARAS3 experiment disfavored the presence of an EDGES-like cosmic signal ~\citep{Singh:2021mxo}.

It is somewhat surprising that many analyses reached different conclusions about the existence and/or depth of the cosmic 21cm profile in EDGES data.
Ideally, any interpretation should use self-consistent forward models with physically-informed priors.
The vast majority of radio excess and gas cooling analysis (see e.g. \citealt{ewall2018modeling,barkana2018possible,Munoz:2018pzp,mirocha2019does,Ewall-Wice:2019may,Reis:2020arr}) instead compute a "pseudo likelihood" using only a {\it summary} of the data, 
typically the center and (or) width of the \citetalias{Bowman:2018yin} profile.
This approach is intrinsically problematic since the \citetalias{Bowman:2018yin} profile was computed assuming a phenomenological "flattened Gaussian" $\Tbb$  and therefore cannot be interpreted directly with a {\it different} (e.g. physical) model for the $\Tbb$.\footnote{A CMB analogy to this common approach would be if one took the best fit EoR history from {\it Planck} obtained assuming a phenomenological tanh shape, $\hat{x}_{\rm HII, tanh}(z)$, and then performed inference using a physical, galaxy-driven model for the EoR history, $x_{\rm HII, gal}(z)$, but constructing a Gaussian likelihood around $\hat{x}_{\rm HII, tanh}(z) - x_{\rm HII, gal}(z)$.  To our knowledge, this has not been done in CMB analyses.}

Here we directly forward model the EDGES sky temperature data,
comparing the Bayesian evidence for a range of foreground models and residual calibration systematics. 
Unlike the analytic and phenomenological models used in \citetalias{Bowman:2018yin}, \citetalias{Sims:2019kro} and~\citealt{Murray:2022uhg}, we build on the cosmological {\tt 21cmFAST} simulation code (e.g. \citealt{Mesinger:2010ne, Murray:2020trn}), including an inhomogeneous, excess radio background sourced from the first, molecular-cooling galaxies that are hosted by $\sim10^5-10^8 M_\odot$ halos.  
These primordial galaxies are expected to be dominated by metal-free, so-called  population III (Pop III) stars and their remnants, which could have very different properties compared to later generations (e.g. \citealt{Woosley:2002zz,Heger:2002by}).  
For simplicity, we refer to molecular-cooling galaxies as "Pop III galaxies" in this work.
Later generations of galaxies form out of Pop III galaxy seeds, and mainly reside in more massive dark matter halos ($>10^8 M_\odot$)  for which atomic cooling is efficient.  Analogously to Pop III galaxies, here we will refer to atomic cooling galaxies as Pop II galaxies, as most of their stellar population is formed out of pre-enriched gas (e.g.~\citealt{Bromm:2003vv, Trenti:2010hs,Salvaterra:2010nb}).

Previously it was shown in~\citealt{mirocha2019does} and \citealt{Reis:2020arr} that in order for Pop II galaxies to reproduce the amplitude of the \citetalias{Bowman:2018yin} feature,
the corresponding radio background would exceed present-day measurements from  the Absolute Radiometer for Cosmology, Astrophysics, and Diffuse Emission 2 (ARCADE2; \citealt{Fixsen:2009xn}) by orders of magnitude.
This tension was resolved in an ad-hoc manner by introducing a phenomenological redshift cut-off parameter, $z_{\rm off}$ (see e.g. ~\citealt{mirocha2019does, Reis:2020arr,Sikder:2022hzk,Sikder:2024osn}),
below which galactic radio emission is quenched.
In contrast, Pop III galaxies are susceptible to feedback from the Lyman-Werner (LW) background (e.g. \citealt{Qin:2020xyh,Munoz:2021psm}) which eventually sterilizes star formation inside them.
As a result,
the radio emission from Pop III galaxies decays naturally once a LW background is established, 
making Pop III galaxies a natural candidate to explain EDGES while being consistent with ARCADE2 measurements.

In this work, we determine whether the EDGES data prefers such a physical model for a high-redshift radio-background excess.  We forward-model the global signal varying Pop III galaxy properties, foregrounds and systematics/calibration residuals.  We vary the complexity of the foreground and systematics models, comparing their Bayesian evidences.  We include complementary data from ARCADE2 and {\it Planck}.  To speed up the inference, we train an emulator of summary observables in our model, providing it as a new option in the public {\tt 21cmEMU}
\footnote{\href{https://github.com/21cmfast/21cmEMU}{https://github.com/21cmfast/21cmEMU}} 
package~\citep{Breitman:2023pcj}.

The paper is organized as follows. 
\Sec{Sec_EDGES_Observation} reviews the EDGES observation, while 
\Sec{Sec_Models} details our model for Pop III radio galaxy.
 We build some physical intuition about our model with an illustrative example in \Sec{Sec_Intuition} and then discuss our model for foreground emission and systematics in \Sec{Sec_FG_and_calibration}.
Our likelihoods and inference methodology are detailed in Secs. \ref{Sec_Likelihods} and \ref{Sec_Inference},
respectively.
\Sec{Sec_Results} shows our results.
Finally we present discussions and caveats in \Sec{Sec_Discussion_caveats} and conclude in \Sec{Sec_Conclusions}.
We assume $\Lambda \r{CDM}$ cosmology with the relevant parameters set by {\it Planck} 2018 results~\citep{Planck:2018vyg}:
$H_0 = 67.66\ \r{kms^{-1}Mpc^{-1}}$,
$\Omega_{\Lambda} = 0.6903$,
$\OmM = 0.3096$,
$\OmC = 0.2607$,
$\OmB = 0.0489$,
$\ln(10^{10}A_\r{s}) = 3.047$,
$n_\r{s} = 0.967$.

\section{EDGES observations}
\label{Sec_EDGES_Observation}

We use the same publicly-available EDGES data as in \citetalias{Bowman:2018yin} (c.f. \citealt{Murray:2022uhg}),
namely the sky temperature spectrum $T_\r{sky}$
\footnote{
\href{http://loco.lab.asu.edu/edges/edges-data-release}{http://loco.lab.asu.edu/edges/edges-data-release}}.
This section gives a brief review of the observation and data processing, for more details, see \citetalias{Bowman:2018yin}.

The data are from the EDGES low-band instruments operating over 50-100 MHz frequencies,
which correspond to $13 \le z \le 27$.
The data were collected between 2016 and 2017 and consist of a total of 138 days of observation, after the initial quality cuts~\citep{Murray:2022uhg}.
The raw data are contaminated by transient sources such as the sun, weather and radio frequency interference~\citep{Tauscher:2020wso},
the calibration and data-analysis pipeline removes and corrects for these effects resulting in $T_\r{sky}$,
which was assumed to consist of the beam-weighted foregrounds (FG) and the  global cosmic  21cm signal~\citep{Tauscher:2020wso}:
\be
T_\r{sky}
=
T_\r{FG}
+
\Tbb.
\label{Eq_B18_Tsky_Model}
\ee

\citetalias{Bowman:2018yin} modeled the foregrounds as a 5-term log-polynomial, motivated by the known properties of the Galactic synchrotron spectrum and the Earth's ionosphere,
\be
\begin{aligned}    
T_\r{FG}
\approx
&
a_0
\left(\frac{\nu}{\nu_\r{c}}\right)^{-2.5}
+
a_1
\left(\frac{\nu}{\nu_\r{c}}\right)^{-2.5}
\ln
\left(\frac{\nu}{\nu_\r{c}}\right)
\\
&
+
a_2
\left(\frac{\nu}{\nu_\r{c}}\right)^{-2.5}
\left[
\ln
\left(\frac{\nu}{\nu_\r{c}}\right)
\right]^2
+
a_3 \left(\frac{\nu}{\nu_\r{c}}\right)^{-4.5}
+
a_4 \left(\frac{\nu}{\nu_\r{c}}\right)^{-2},
\end{aligned}
\ee
here $\nu$ is the observing frequency,
$a_\r{n}$ are fitting coefficients,
and $v_\r{c} = 75$ MHz is the central frequency of the observing band.

The cosmic 21~cm signal  in 
\Eq{Eq_B18_Tsky_Model} was modelled as a flattened-Gaussian:
\be
\Tbb
=
-A
\left[
\frac{
1-\exp
\left(
-\tau \r{e}^B
\right)
}
{
1-\exp
\left(
-\tau
\right)
}
\right],
\ee
with,
\be
B
=
\frac{4 (\nu - \nu_0)^2}
{w^2}
\ln
\left[
-\frac{1}{\tau}
\ln
\left(
\frac{1+\r{e}^{-\tau}}
{2}
\right)
\right],
\ee
where $A$, $\nu_0$, $w$ and $\tau$ are model parameters representing the depth,
central frequency,
full-width at half-maximum and flatness of the signal, respectively.
Inference using this model resulted in a noise-like residual with an RMS (root mean square) of 25mK, and 
constrained the $\Tbb$ parameters to $A=0.5^{+0.5}_{-0.2}\r{K}$,
$\nu_0 = 78\pm\r{MHz}$,
$w = 19^{+4}_{-3}\r{MHz}$ and $\tau = 7^{+3}_{-3}$,
where the bounds represents 99\% credible intervals (C.I.s).

In this work, we forward-model the calibrated brightness temperature data, $T_\r{sky}$, using our own physical model for the 21cm signal, $\Tbb$, as well as different basis sets for foregrounds, $T_\r{FG}$,  and possible calibration residuals, $T_\r{cal}$ (c.f. equation \ref{eq:skytemp}).
We describe our models for each in turn below.

\section{Enhanced 21cm absorption via radio-loud, molecular-cooling galaxies}
\label{Sec_Models}

We extend the public {\tt 21cmFAST} code~\citep{Mesinger:2010ne, Murray:2020trn} to include an inhomogeneous radio background from galaxies.
Starting from cosmological initial conditions,
{\tt 21cmFAST} simulates 3D lightcones of density,
star formation and various radiation fields to be used for computing the thermal and ionisation evolution of the IGM.  Here we summarize some of the salient points of this procedure, highlighting the novelties of our model; readers are encouraged to consult the above references for more details on the {\tt 21cmFAST} code.

We make the ansatz that the very first molecularly-cooled Pop III galaxies were responsible for the radio excess putatively observed by EDGES.
The star formation in Pop III galaxies is transient as they are sterilized by the build-up of a Lyman-Werner (LW) background and feedback from photoheating ~\citep{Qin:2020xyh,Munoz:2021psm}.
Therefore they provide a physical way to establish an early radio background,
without having to introduce a "turn-off" redshift (see e.g.~\citealt{mirocha2019does, Reis:2020arr}) so as not to overproduce the present-day radio background.  Below we discuss our star formation prescription, and how we model the associated radiation backgrounds using semi-empirical relations.

\subsection{Star formation}
\label{SubSec_SFRD}

We assume that stars form in both atomic  and molecular cooling halos, and that these can have different properties.
Made from the pristine unpolluted metal-free gas,
the first (Pop III) stars are expected to form in small molecular-cooling galaxies at $z \sim 20-30$~\citep{Qin:2020xyh,Qin:2020pdx,Munoz:2021psm},
whereas the second (Pop II) generation of stars form from the gas polluted by metallic remnants of Pop III stars and reside in galaxies that accrete gas through atomic-cooling.
Their comoving star formation rate density (SFRD) is linked to the halo mass function according to:
\be
\r{SFRD}_\r{s}
=
\int
\rd M_\r{h}
\ 
\dot{M}_{\star,\r{s}}
f_{\r{duty,s}}
\frac{\rd n}
{\rd M_\r{h}} \left(M_\r{h},z|R,\delta_{\r{R}}\right).
\label{Eq_SFRD}
\ee
Hereafter we use the subscript $\r{s}$ to denote Pop II (s=II) and Pop III (s=III) galaxies,
$M_\r{h}$ is halo mass,
and $\rd n / \rd M_\r{h} \left(M_\r{h},z|R,\delta_{\r{R}}\right)$ is the conditional halo mass function,
which gives the differential halo number density for a region of scale $R$ and corresponding overdensity $\delta_\r{R}$.
Equation (\ref{Eq_SFRD}) is evaluated in different patches of the simulation box, with the density field modulating (i.e. conditioning; \citealt{Lacey:1993iv,Somerville:1997df,Cooray:2002dia}) the halo mass function and thus sourcing spatial variations in the SFRD.

The stellar mass of a galaxy is assumed to build up over some fraction, $\eta$, of the Hubble time, $H^{-1}(z)$,
where $H(z)$ is the Hubble parameter.
Thus the star formation rate $\dot{M}_{\star,\r{s}}$ is related to the stellar mass $M_{\star,\r{s}}$ via,
\be
\dot{M}_{\star,\r{s}}
=
M_{\star,\r{s}}
H(z)/\eta,
\ee
where $\eta$ is a free parameter between zero and unity,
and
\be
M_{\star,\r{s}}
=
f_{\star,\r{s}}
\left(
\frac{\OmB}{\OmM}
\right)
M_\r{h}.
\label{Eq_Stellar_Mass}
\ee
here $f_{\star,\r{s}}$ is the fraction of baryons converted into stars,
\be
f_{\star,\r{II}}
=
\min
\left[
f_{\star, 10}
\left(
\frac{M_\r{h}}
{10^{10} M_\odot}
\right)^{\alpha_{\star,\r{II}}},\ 
1
\right],
\ee
\be
f_{\star,\r{III}}
=
\min
\left[
f_{\star, 7}
\left(
\frac{M_\r{h}}
{10^{7} M_\odot}
\right)^{\alpha_{\star,\r{III}}},\ 
1
\right].
\label{Eq_Fstar_III}
\ee

Finally in \Eq{Eq_SFRD},
the duty cycle $f_\r{duty,s}$ describes the star formation efficiency inside the halo,
\be
f_{\r{duty, II}} = \exp\left( -M_{{\r{turn,II}}}/M_\r{h}\right),
\ee
\be
f_{\r{duty, III}} = \exp\left( -M_{{\r{turn,III}}}/M_\r{h}\right)
\exp\left( -M_\r{h}/M_{\r{atom}}\right).
\label{Eq_fduty_III}
\ee
Here the turnover mass $M_\r{turn,s}$ characterizes the mass scale below which star formation is strongly suppressed.
$M_\r{turn,s}$ is determined by various feedback mechanisms and can be expressed as,
\be
M_\r{turn,II}
=
\max
(M_\r{crit}^\r{ion},
M_\r{atom}),
\label{Eq_Mturn_II}
\ee
\be
M_\r{turn,III}
=
\max
(
M_\r{crit}^\r{ion},
M_\r{mol}
),
\label{Eq_Mturn_III}
\ee
where $M_\r{crit}^\r{ion}$ and $M_\r{atom}$ describe photoheating feedback and the atomic cooling threshold (corresponding to a halo virial temperature of $10^4$ K; see~\citealt{Sobacchi:2013ww,Qin:2020xyh,Qin:2020pdx}).

Efficient star formation inside Pop III galaxies also depends on the inhomogeneous LW background and the local DM-baryon relative velocity.  Following \citealt{Munoz:2021psm}, we take,
\be
M_\r{mol}
=
M_0(z)
f_{v_{\r{cb}}}
f_{\r{LW}},
\label{Eq_mmol}
\ee
where $f_{v_{\r{cb}}}$ is determined by an empirical fit to hydrodymanical simulations  \citep{Kulkarni:2020ovu,Schauer:2020gvx}, and $M_0(z) = 3.3 \times 10^7 (1+z)^{-3/2}M_\odot$.
$f_\r{LW}$ describes LW feedback,
which will be detailed in the next subsection.

\subsection{Inhomogeneous cosmic radiation fields}

We briefly summarize how we compute the LW,
X-ray, ionizing UV and radio radiation fields,
which are relevant for the 21cm signal as well as for photo-heating (ionizing) and photo-dissociating (LW) feedback on star formation.
We encourage interested readers to see ~\citealt{Munoz:2021psm} and ~\citealt{Qin:2020xyh} for more details.

\subsubsection{Lyman Werner radiation}
Lyman Werner radiation (whose photons have energies in the range 11.2 - 13.6 eV) can photodissociate $\r{H}_2$ molecules 
through the Solomon process. 
Pop III galaxies exposed to a high LW flux will therefore have difficulty accreting gas through the $\r{H}_2$ cooling channel, which eventually sterilizes their star formation (e.g. \citealt{Tegmark:1996yt,Bromm:2003vv}).
It was shown in~\citealt{Machacek:2000us} that the minimum mass of star forming halos,
approximated here by the turnover mass $M_\r{turn,III}$ in \Eq{Eq_Mturn_III},
has a power-law dependency on LW intensity.  
Motivated by  results from hydrodynamic simulations \citep{Kulkarni:2020ovu,Schauer:2020gvx},
{\tt 21cmFAST} adopts the analytic form in~\citealt{Visbal:2014fta} to parameterize the suppression factor $f_\r{LW}$ in \Eq{Eq_mmol},
\be
f_{\r{LW}}
=
1+A_\r{LW}
J_{\r{LW}}^{\beta_\r{LW}},
\label{Eq_LW}
\ee
where $A_\r{LW}$ and $\beta_\r{LW}$ are free model parameters,
and $J_{\r{LW}}$ is the LW intensity in $10^{-21} \r{erg s^{-1} cm^{-2} Hz^{-1} sr^{-1}}$, at a redshift $z$, and cell position ${\bf x}$,
\be
J_{\r{LW}}(z, {\bf x})
=
\frac{c(1+z)^3}{4 \pi}
\int_z^{\infty}
\frac{\rd z'}{(1+z')H}
\epsilon_\r{LW}
\r{e}^{-\tau_{\r{LW}}}.
\ee
Here the optical depth $\tau_{\r{LW}}$ accounts for resonance attenuation,
and $\epsilon_\r{LW}$ is the LW emissivity from both Pop II and Pop III galaxies~\citep{Qin:2020xyh}.

\subsubsection{X-ray emission}

X-rays that ionize and heat the IGM are primarily emitted by high mass X-ray binaries (HMXBs), 
which are relatively short-lived and so trace the star formation rate of the host galaxy (e.g.~\citealt{Fragos:2012vf}).
Therefore the rest-frame, specific X-ray emissivity $\epsilon_\r{x}$ is taken to be proportional to the SFRD (consistent with observations of low-redshift star forming galaxies; e.g. \citealt{Lehmer:2020rqq}):
\be
\epsilon_{\r{x,s}}(z, {\bf x}, \nu)
=
L_{\r{x,s}}/\r{SFR}
\times
\r{SFRD}_\r{s}.
\ee
Here $L_{\r{x,s}}/\r{SFR}$ is the specific X-ray luminosity per star formation rate (SFR), which is assumed to follow a power-law frequency spectrum,
\be
L_{\r{x,s}}/\r{SFR}
\propto
\nu^{\alpha_\r{x}},
\ee
and $\alpha_\r{x}$ is the power-law index.
We define the normalization parameter of $L_{\r{x},s}/\r{SFR}$ as,
\be
\mathcal{L}_\r{x,s}
\equiv
\int_{E_0}^{2\,\r{keV}}
\rd E\ 
L_\r{x,s}/\r{SFR},
\label{Eq_LX}
\ee
where $E_0$ is the energy threshold below which photons are absorbed by the host galaxy (e.g.~\citealt{Das:2017fys}).

\subsubsection{Ionizing UV}

Reionization of the IGM is assumed to be driven by ionizing UV radiation from massive stars.
Following the excursion-set formalism of \citealt{Furlanetto:2004nh},
{\tt 21cmFAST} identifies a simulation cell as ionized if the cumulative number of ionizing photons per baryon $\bar{n}_\r{ion}$ satisfies the following condition over a spherical region with any given radius $R$,
\be
\bar{n}_\r{ion}
\geq
(1+\bar{n}_\r{rec})
(1-\bar{x}_\r{e}),
\ee
where $\bar{n}_\r{rec}$ is the cumulative number of recombinations per baryon (see~\citealt{Sobacchi:2014rua,Qin:2020xyh}),
$\bar{x}_\r{e}$ is the average ionization fraction induced by X-rays,
and $\bar{n}_\r{ion}$ is calculated by,
\be
\bar{n}_\r{ion}
=
\frac{1}{\rho_\r{b}}
\sum_\r{s}
\int \rd M_\r{h}
\phi_\r{s}
M_{\star,\r{s}}
n_{\gamma,\r{s}}
f_\r{esc,s},
\ee
where $\rho_\r{b}$ is the averaged baryon density within a spherical region of radius $R$,
$\phi_\r{s} = f_\r{duty,s} \rd n/\rd M_\r{h}$,
the stellar mass $M_{\star,s}$ is given in \Eq{Eq_Stellar_Mass},
$n_{\gamma,\r{s}}$ is the number of ionizing photons emitted per stellar baryon,
for which we adopt $5\times10^3$ for Pop II and $5\times10^4$ for Pop III.
Finally $f_\r{esc,s}$ is the fraction of ionizing UV photons that escape the host galaxies, normalized at the same characteristic halo mass values as the SFR efficiency:
\be
f_{\r{esc, II}}
=
\min
\left[
f_{\r{esc}, 10}
\left(
\frac{M_\r{h}}
{10^{10} M_\odot}
\right)^{\alpha_\r{esc,II}},\ 
1
\right],
\ee

\be
f_{\r{esc, III}}
=
\min
\left[
f_{\r{esc}, 7}
\left(
\frac{M_\r{h}}
{10^{7} M_\odot}
\right)^{\alpha_\r{esc,III}},\ 
1
\right].
\label{Eq_Fesc_III}
\ee

\subsubsection{Radio emission}

Radio emission in star-forming galaxies is primarily sourced by synchrotron radiation associated with the end products of short-lived massive stars, resulting in well-established linear scalings between radio luminosity and SFR (e.g.~\citealt{condon2002radio, Heesen:2014vga,gurkan2018lofar}).  Analogously to the X-ray emissivity, we therefore model the rest-frame comoving specific radio emissivity $\epsilon_\r{R}$ as proportional to the SFRD,
\be
\begin{aligned}
\epsilon_{\r{R,s}} (z, {\bf x}, \nu)
&=
f_\r{R,s}
10^{22}
\left(
\frac{\nu}{0.15 \r{GHz}}
\right)^{-\alpha_\r{R,s}}
\\&
\times \frac{\r{SFRD_s}}{M_{\odot} {\r{yr^{-1} Mpc^{-3}}}}
\ 
\r{w Hz^{-1} Mpc^{-3}},
\end{aligned}
\label{Eq_EMS_Radio}
\ee
where $f_\r{R,s}$ is the radio emission efficiency,
which is roughly unity for galaxies observed today ~\citep{mirocha2019does, Reis:2020arr}, but which here we allow to be orders of magnitude larger for Pop III galaxies.  The frequency spectrum in the above equation is taken to be a power-law with a spectral index of $\alpha_\r{R,s}$ (discussed further below). 

The relevant radio brightness temperature at 21~cm can be computed as,
\be
T_{\r{radio}}
(z,\nu_{21})
=
\frac{c^3(1+z)^3}
{8 \pi k_\r{B} \nu^2_{21}}
\int_z^{\infty}
\frac{\rd z'}
{(1+z')H}
\epsilon_\r{R} (\nu'_{21},z'),
\label{Eq_T_Radio_Integration}
\ee
where $\nu_{21}=1.42 \r{GHz}$,
$z'$ is the emission redshift,
$\nu_{21}' = \nu_{21}(1+z')/(1+z)$ is the emission frequency,
$c$ is the speed of light in vacuum,
$k_\r{B}$ is the Boltzmann constant.

The ARCADE2 experiment~\citep{Fixsen:2009xn} and measurements from earlier instruments~\citep{haslam1981408,reich1986radio,Roger:1999jy,maeda199945} showed that the current radio excess background temperature in the $[0.022, 90]\r{GHz}$ frequency range can be well described by a power-law,
\be
T(\nu)
=
T_\r{r}
\left(
\frac{\nu}
{\r{GHz}}
\right)^{\beta},
\label{Eq_ARCADE_z0}
\ee
where,
\be
T_\r{r}
=
1.19\pm0.14 \r{K}, 
\beta = -2.62\pm0.04.
\label{Eq_ARCADE_model_conatraints}
\ee
Subsequent measurements from the Long Wavelength Array (LWA) ~\citep{Dowell:2018mdb}   yielded similar results.
It can be inferred from Eqs. (\ref{Eq_EMS_Radio}) and (\ref{Eq_T_Radio_Integration}) that at arbitrary redshifts,
our $T_{\r{radio}}$ is proportional to $\nu^{-(2+\alpha_{\r{R},s})}$.
Thus we fix $\alpha_{\r{R},s}$ to $0.62$ hereafter,
which gives a spectral shape of $\Tr \propto \nu^{-2.62}$ in agreement with ARCADE2.

\subsection{21cm signal}

The cosmological 21cm signal is typically denoted by its brightness temperature, $\Tb$, which can be approximated as~(e.g. \citealt{Furlanetto:2004zw,Pritchard:2011xb}),
\bea
\Tb (z, {\bf x})
& \simeq &
27 
\xH
\left(
1 + \delta_{\rm b}
\right)
\left(
\frac{H}
{
\rd v_\r{r}
/
\rd r
+
H
}
\right)
\left(
1 - 
\frac{T_\r{R}}{T_\r{s}}
\right) \nonumber
\\
&\times &
\left(
\frac{1+z}{10}
\frac{0.15}
{\Omega_\r{m}h^2}
\right)^{1/2}
\left(
\frac{\Omega_\r{b}h^2}
{0.023}
\right)
\r{mK}.
\label{Eq_T21}
\eea
Here $\xH$ and $\delta_\r{b}$ denote the hydrogen neutral fraction and baryon over-density,
respectively,
$\rd v_\r{r} / \rd r$ is the radial gradient of velocity field,
$\OmM$ and $\OmB$ are fractional densities in matter and baryons,
respectively,
and
$h$ is the Hubble constant in $100\ \r{km\ s^{-1}\ Mpc^{-1}}$.
In the presence of radio galaxies,
the background radio temperature $T_\r{R}$ takes the form,
\be
T_\r{R}
=
T_\r{CMB}
+
T_\r{radio},
\ee
where $T_\r{CMB} = 2.728(1+z)\r{K}$ is the CMB temperature,
and $T_\r{radio}$ is computed following \Eq{Eq_T_Radio_Integration}.

Finally in \Eq{Eq_T21} the gas spin temperature $T_\r{S}$ 
parameterizes the ratio of number densities of hydrogen in the spin triplet ($n_1$) and spin singlet ($n_0$) states:
$n_1/n_0 = 3 \exp(-0.068\r{K}/T_\r{S})$,
and it is coupled to both $T_\r{R}$ and the gas kinetic temperature $T_\r{K}$ by~\citep{Pritchard:2011xb,mirocha2019does},
\be
T^{-1}_\r{S}
=
\frac{
T^{-1}_\r{R}
+
x_\alpha
T^{-1}_\alpha
+
x_\r{c}
T^{-1}_\r{K}
}
{1 + x_\alpha + x_\r{c}},
\label{Eq_Ts}
\ee
where $T_\alpha \simeq T_\r{K}$ is the color temperature~\citep{Hirata:2005mz}, and 
$x_\alpha$ and $x_\r{c}$ are coefficients for collisional and Wouthuysen-Field coupling (see~\citealt{Pritchard:2011xb}).

\section{Building physical intuition}
\label{Sec_Intuition}

\subsection{An illustrative example}

\begin{figure*}[ht]
\centering
\includegraphics[width=20cm]{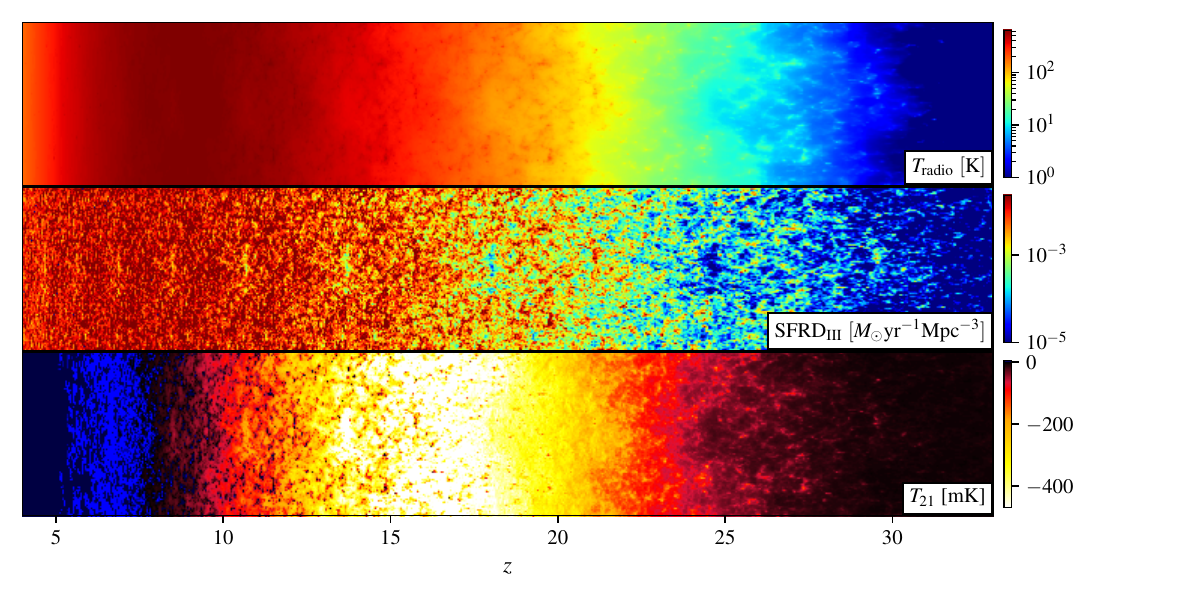}
\caption{Slices through the  $\Tr$ (top), Pop III SFRD (middle) and $\Tb$ (bottom) lightcones for our illustrative model (see text for details).  The simulation presented here has a box size of $500^3$ $\r{Mpc^{3}}$ with a resolution of $250^3$.  In order to accentuate the cosmic dawn era, we plot the horizontal axes linearly in redshift (as opposed to linearly in comoving scale).
}
\label{Fig_LightCones}
\end{figure*}

\begin{figure*}[ht] 
\centering
\includegraphics[width=18cm]{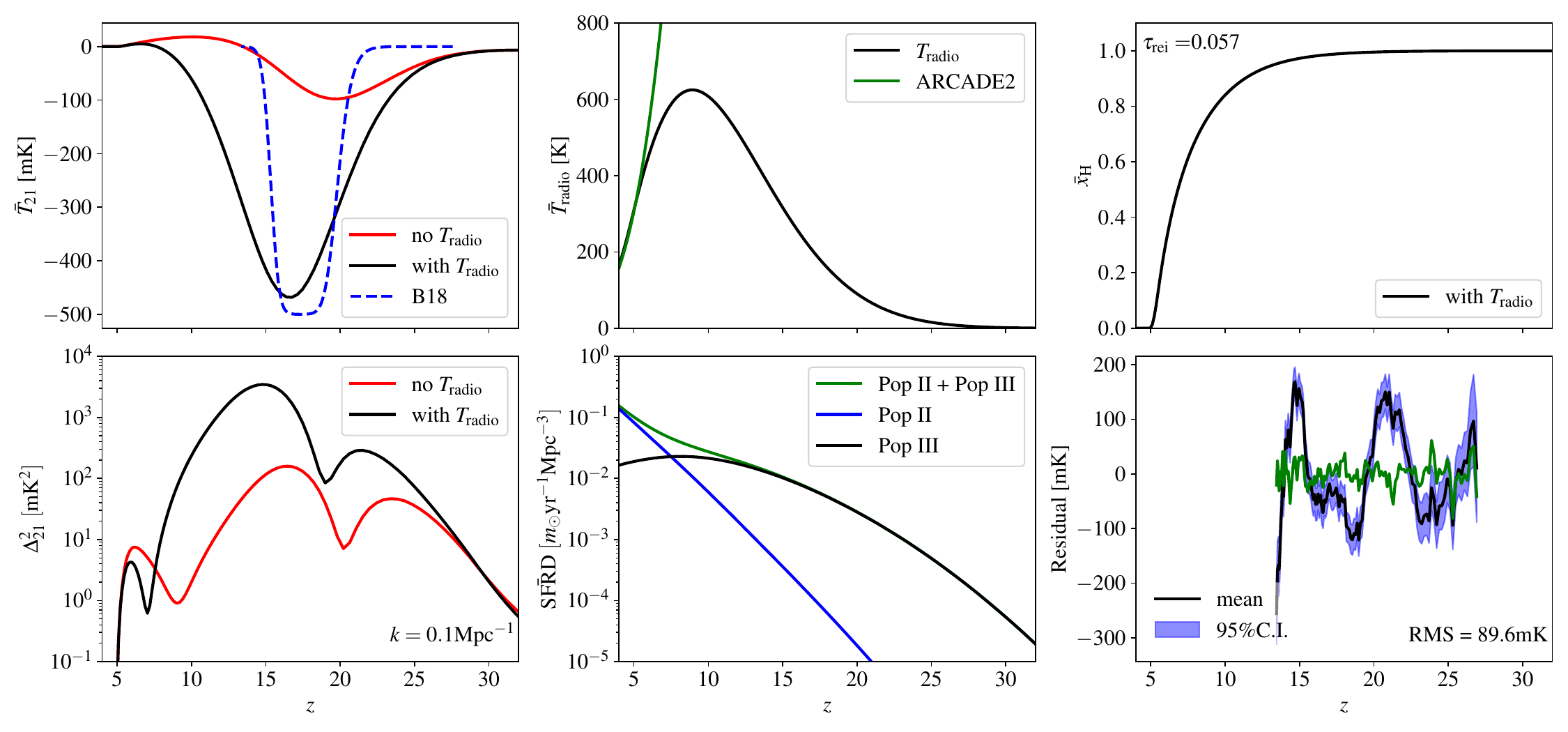}
\caption{
Summaries for the simulation shown in \Fig{Fig_LightCones}.
In the top panels,
from left to right,
the black solid lines show global average values for $\Tb$, $T_\r{radio}$ and $\xH$, respectively.
The blue dashed line in the $\Tbb$ panel shows the MAP flattened Gaussian profile recovered by ~\citetalias{Bowman:2018yin}, while the red curve corresponds to the same astrophysical model as shown with the black curve but without a radio background excess.  The
radio background observed by ARCADE2 (see Eqs.(\ref{Eq_ARCADE_z0}, \ref{Eq_T_ARCDDE_v21})) is indicated in the $\bar{T}_\r{radio}$ panel with the green solid line.
The lower left panel shows the 21cm power spectrum at $k=0.1\r{Mpc^{-1}}$,
with black and red solid lines corresponding to the cases with and without a radio excess, respectively.
We show the global star formation rate density in the lower middle panel,
where the blue, black and green solid lines represent contributions from Pop II galaxies, Pop III galaxies and their sum, respectively.
In the lower right panel,
we present the brightness temperature residuals (mean and 95\% C.I.) after performing inference on EDGES data with a 5-term FG model and fixing the cosmic signal to the $\Tbb$ profile shown with the black solid line in the top left panel.  There is an obvious, unaccounted for signal remaining in these residuals.  For comparison, we also show in this panel the noise-like mean from our highest evidence model (green solid line; see \Sec{Sec_Results} for details).
}
\label{Fig_Example_1D}
\end{figure*}

Before moving on to the inference results, we first build some physical intuition about our model.  Here and throughout, we only concern ourselves with the properties of Pop III galaxies.  As mentioned earlier, remnants of Pop III star formation might be exotic enough to source a radio background in excess of the CMB at $z\gtrsim17$, and their eventual sterilization through LW and photo-heating feedback provides a physically-motivated way of avoiding the limits set by measurements of the present-day radio background.

We vary five Pop III galaxy properties that regulate their star formation, radiative efficiencies and the strength of LW feedback, as discussed in the previous section: $\left[ f_\r{R,III},
f_{\star, 7}, f_{\r{esc}, 7}, A_\r{LW}, \mathcal{L}_\r{X,III}\right]$.  The remaining parameters, such as those governing Pop II galaxy properties, we set to the default values in the Evolution of 21cm Structure (EoS) 2021 release \citep{Munoz:2021psm}.  Pop II galaxies are dominant at lower redshifts ($z\lesssim10$; c.f. Fig. \ref{Fig_Example_1D}), and indeed the EoS 2021 galaxy model is consistent with existing observations at those epochs, including {\it Hubble } UV luminosity functions~\citep{Bouwens:2014fua,Bouwens:2015vha,Oesch_2018}, 
the CMB optical depth~\citep{Planck:2016mks,Planck:2018vyg} and additional constraints on reionization timing (e.g. \citealt{McGreer:2014qwa}).  Similarly, the radio efficiency of Pop II galaxies is set to zero, though we confirm that the cosmic signal is insensitive to this choice (e.g. setting $f_\r{R,II}=1$ results in a radio background that is $\sim$ 5 orders of magnitude lower than the CMB at the redshifts of interest).

We first vary our Pop III galaxy parameters in order to find a model that is seemingly consistent with both EDGES and ARCADE2.  For this illustrative case, we take the parameter combination $f_\r{R,III} = 1.14 \times 10^3$,
$f_{\star, 7} = 0.06$,
$f_{\r{esc}, 7} = 1.71 \times 10^{-3}$,
$A_\r{LW} = 3.80 \times 10^{-3}$ and
$\mathcal{L}_\r{X,III} = 1.12 \times 10^{40} \r{erg s^{-1} M_\odot^{-1} yr}$.
\Fig{Fig_LightCones} shows the corresponding lightcone evolutions of $\Tb$, $\Tr$ and the SFRD, evaluated on the scale of the simulation cells.
As expected, the fluctuations in the radio background trace the underlying SFRD, though they are smoothed by photon propagation.   Because in this model the first Pop III galaxies are three orders of magnitude more radio luminous per unit SFR compared to local ones, the 21cm brightness temperature has a minimum of $\lesssim -450$mK, well in excess of what "standard" models can produce.  From the top panel, we see  that the mean intensity of the radio background {\it decreases} below $z \lesssim 10$.  This demonstrates that the Pop III galaxies can indeed source a radio background excess during the cosmic dawn which then naturally fades towards lower redshifts.

We further quantify the above claims in \Fig{Fig_Example_1D}.
We show in the top panels the evolution of $\Tbb$, $\bar{T}_\r{radio}$ and $\bar{x}_\r{H}$, respectively,
where here and below we use $\bar{x}$ to denote globally-averaged values of the physical quantity $x$.  We see that indeed the 21cm global signal in this model is comparable in amplitude and timing to the MAP (Maximum a Posteriori) flattened Gaussian model recovered by EDGES ({\it dashed blue line in the top left panel}).  It achieves this without exceeding the ARCADE2 radio background measurements ({\it middle panel}) and having an reionisation history that is consistent with current constraints ({\it right panel}).  We recall that by construction, the model is consistent with {\it Hubble} observations of UV LFs at $z \gtrsim 6$, as these only constrain relatively bright Pop II galaxies whose values we fix to those in \citealt{Munoz:2021psm}.

The lower left panel of \Fig{Fig_Example_1D} shows the evolution of the large-scale ($k=0.1$ Mpc$^{-1}$) 21cm power spectrum (PS) $\Delta^2_{21}$, defined as
\be
\left<
\tilde{T}_{21}(k, z)
\tilde{T}^*_{21}(k', z)
\right>
\equiv
(2 \pi)^3 \delta_\r{D}
(k - k')
\frac{2 \pi^2}{k^3}
~
\Delta^2_{21}(z,k)
,
\ee
where $\tilde{T}_{21}$ indicates the 3D Fourier transform of $\Tb$.\footnote{
All our 21cm power spectrum are computed from {\tt 21cmFAST} simulations using the {\tt powerbox} package~\citep{Murray2018} ,
which is available at \href{https://github.com/steven-murray/powerbox}{https://github.com/steven-murray/powerbox}.}
The black curve corresponds to the fiducial model discussed in this section, while the red curve corresponds to the same model but assuming no radio background excess (i.e. $f_\r{R,III}=0$).
In both cases, the large-scale power exhibits the usual three peak evolution, marking (from left to right) the epochs of reionization (EoR), X-ray heating (epoch of heating, EoH) and Lyman-Alpha pumping~\citep{Pritchard:2006sq,Mesinger:2010ne,Lopez-Honorez:2016sur}.
However, the addition of a population of radio loud Pop III galaxies enhances the peak amplitude of the large-scale power during the EoH by over an order of magnitude (see also~\citealt{Fialkov:2019vnb,Reis:2020arr}). 

In the bottom middle panel, we show the relative contributions to the total SFRD ({\it green curve}), provided by Pop II galaxies ({\it blue curve}) and Pop III galaxies ({\it black curve}).  We see explicitly that Pop III galaxies in this model dominate the SFRD during the CD, at $z\gtrsim10$.  At lower redshifts, their SFRD starts to decrease due to a combination of LW feedback, photo-heating feedback and the evolution of the halo mass function (c.f. \citealt{Munoz:2021psm}).

Based on the previously-mentioned results, it would seem this Pop III radio background model can roughly reproduce the EDGES data, while being consistent with complementary observations.  If we define a "pseudo-likelihood" using the timing and depth of the flattened Gaussian recovered by EDGES, we might conclude this model provides a good description of the data.  However, our cosmic signal is not a flattened Gaussian, and we should define a likelihood directly in data space (i.e. the observed sky temperature).  

To check how well 
this model actually reproduces the data, we perform an inference in which we fix the cosmic signal (black solid curve in the top left panel), and sample a 5-term log-polynomial foreground (as in \citetalias{Bowman:2018yin}) using the likelihood  defined by \Eq{Eq_LogLike_EDGES} (the details of our inference procedure will be discussed in Sections \ref{Sec_FG_and_calibration} and \ref{Sec_Likelihods}). In the bottom right panel of \Fig{Fig_Example_1D} we plot the mean and 95\% C.I. of the resulting residual signal (i.e. data minus model). 
If the model were a complete description of the data, these residuals should just be zero mean Gaussian noise.
However, there is clear structure in the residuals,
and this is especially obvious when compared with the residual for our highest evidence model (green solid line, to be detailed in \Sec{Sec_Results}).
Thus, despite the apparent agreement with the \citetalias{Bowman:2018yin} flattened Gaussian in terms of $\Tbb$ depth and timing,
the model {\it does not} explain the data.
We discuss this in more detail in Appendix \ref{Sec_Pseudo_Likelihood} where we further quantify the dangers inherent in using a "pseudo-likelihood" defined on a flattened Gaussian summary of the data.

\subsection{Impact of galaxy parameters}
\label{SubSec_Param_Impacts}

\begin{figure*}[ht]
\centering
\includegraphics[width=18cm]{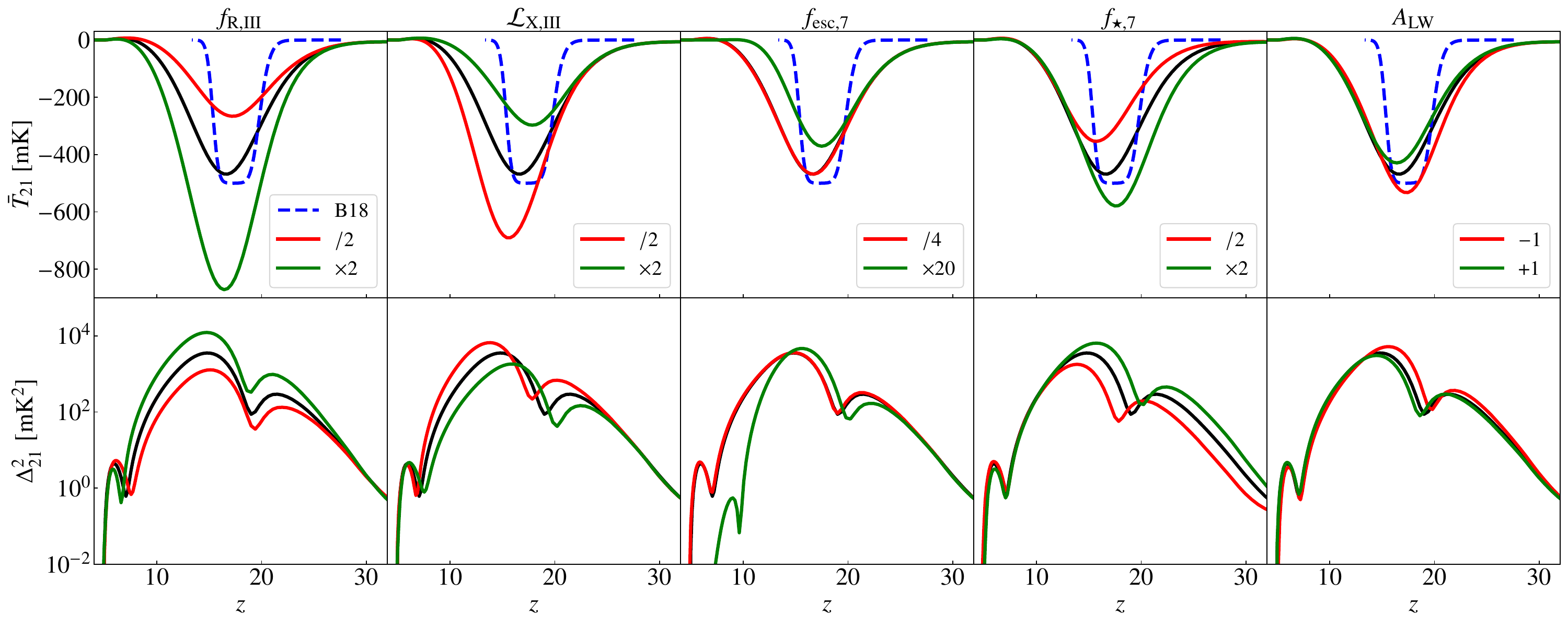}
\caption{
The effects of varying Pop III galaxy parameters on the 21cm global signal (top) and power spectrum (bottom, at $k = 0.1 \r{Mpc^{-1}}$).
From left to right,
we vary $f_\r{R,III}$, $\mathcal{L}_\r{X,III}$, $f_\r{esc,7}$, $f_{\star, 7}$ and $A_\r{LW}$, respectively.  The black solid lines correspond to the fiducial values used to compute  \Fig{Fig_LightCones}, while the green / red solid lines show the result from increasing / decreasing each parameter while keeping the others fixed.
The global 21cm signal reported in~\citetalias{Bowman:2018yin} is shown in the top panels with blue dashed lines.
}
\label{Fig_Trend}
\end{figure*}

In \Fig{Fig_Trend} we visualize how essential aspects of Pop III astrophysics impact the 21cm global signal  and power spectrum.
The black curve corresponds to the fiducial parameter combination discussed in the previous section, while the red/green curves illustrate how each observable changes when decreasing/increasing one parameter at a time in each column.
The variations in $\Delta^2_{21}$ largely follow those in $\Tbb$,
i.e. an enhanced global signal provides an increased dynamic range resulting in more power on large spatial spaces.
Below we summarize our Pop III galaxy parameters and their qualitative impact:

(I) $f_\r{R,III}$ - the radio luminosity to star formation rate for Pop III galaxies, normalized to values seen in local galaxies.  
Efficiencies of $f_\r{R,III} \gtrsim 10^2-10^3$ drive a radio background during the CD that exceeds that of the CMB.
Increasing $f_\r{R,III}$ in this regime deepens the absorption trough in the global signal and drives a corresponding increase in the power spectrum.
The spin temperature is also coupled to $\Tr$ via \Eq{Eq_Ts},
therefore a higher $f_\r{R,III}$ delays the coupling between $T_\r{S}$ and $T_\r{k}$ and thereby shifts the first peak of $\Delta^2_{21}$ to lower redshifts.

(II) $\LXIII$ - the bolometric X-ray luminosity per SFR of Pop III galaxies.
Increasing $\LXIII$ shifts the EoH to earlier times.  Therefore the gas does not have as much time to cool before being heated, and the corresponding global absorption trough and peak in the power spectrum are reduced and shifted to earlier times.

(III) $f_\r{esc,7}$ - the escape fraction of ionizing photons from Pop III galaxies, normalized to halos with a mass of $10^7 M_\odot$.
Increasing $f_\r{esc,7}$ increases the contribution of Pop III galaxies to the EoR.  Shifting the EoR to earlier times reduces the depth of the absorption trough in the global signal.  The corresponding overlap of the EoR and EoH dramatically reduces the late time PS, as the ionization-temperature cross terms become important (c.f. \citealt{Mesinger:2013nua}).

(IV) $f_\r{\star,7}$ - the fraction of galactic gas in stars for Pop II galaxies, normalized to halos with a mass of $10^7 M_\odot$.  Increasing $f_\r{\star,7}$ increases star formation in early sources, shifting all cosmic epochs to earlier times.  Because the radio background excess also scales with the star formation rate, the absorption trough deepens with increasing $f_\r{\star,7}$.

(V) $A_\r{LW}$ - the efficiency of LW feedback on star formation.  Increasing $A_\r{LW}$ makes it easier for a LW background to quench star formation in Pop III galaxies, which has a qualitatively similar impact as decreasing $f_\r{\star,7}$.

\section{Foreground and systematics nuisance parameters}
\label{Sec_FG_and_calibration}

The physically-motivated foreground model used in \citetalias{Bowman:2018yin} was found to lack the flexibility required to capture the beam-weighted foregrounds of the observations (\citetalias{Sims:2019kro}). 
In this work, 
we use a more flexible foreground model, 
as well as an extra term designed to mimic residual structure left over from calibration. 
We summarize them in details below.

\subsection{Foreground model}

The main contaminants of the cosmic 21cm signal come from  galactic foregrounds,
which can be well parameterized by a power-law syncrotron spectrum,
and the Earth's ionosphere~\citep{Hills:2018vyr,bowman2018reply,Sims:2019kro}.
We model their overall contribution using a log-polynomial functional form following~\citealt{Murray:2022uhg},
\be
T_\r{FG}
=
\left(
\frac{\nu}
{75 {\rm{MHz}}}
\right)^{-2.5}
\sum_{i=0}^{N_\r{FG} - 1}
p_i
\left[
\ln
\left(
\frac{\nu}
{
{75 {\rm{MHz}}}
}
\right)
\right]^{i},
\label{Eq_TFG}
\ee
where $\nu$ denotes frequency,
$N_\r{FG}$ is the total number of FG terms, and
$p_i$ are the polynomial coefficients.  We vary the number of foreground terms when maximizing the Bayesian evidence below.

\subsection{Calibration residuals}

Despite the monumental task undertaken by the EDGES team to understand their data pipeline, 
possible imperfections in the calibration process can result in residual systematics~\citep{Hills:2018vyr,Singh:2019gsv,Sims:2019kro}.
As these systematics are "unknown unknowns", 
it is difficult to characterize their contribution via some basis and associated priors.
Here we use the parametrization suggested by \citetalias{Sims:2019kro}.  
Specifically,  
we account for possible calibration systematics using a power-law damped sinusoid model,
\be
T_\r{cal}
=
\left(
\frac{\nu}
{75 {\rm{MHz}}}
\right)^{-2.5}
\left[
a_0
\sin( 2 \pi \nu / P)
+
a_1
\cos( 2 \pi \nu / P)
\right],
\label{Eq_Tcal}
\ee
where $a_0$ and $a_1$ are amplitudes in Kelvin for the sine and cosine components,
and $P$ is the period in MHz.
As noted in \citetalias{Sims:2019kro},
this model of calibration residuals is reasonably well-motivated physically, 
and corresponds to a spectrally-structured gain error in the calibration,
which is weighted by the foreground spectrum. 
Such systematics could in principle be generated by imperfect measurements of the reflection coefficients of the amplifier~\citep{Murray:2022uhg}.

\section{Likelihoods}
\label{Sec_Likelihods}

In addition to EDGES, 
we also use complementary observations from ARCADE2 and {\it Planck},
which allow us to disfavor models that explain the EDGES result by either reionizing the Universe too early or producing a radio background in excess of currently observed limits.  
Thus our final log-likelihood takes the form,
\be
\ln \mathcal{L}
=
\ln \mathcal{L}_{\r{EDGES}}
+
\ln \mathcal{L}_\r{ARCADE2}
+
\ln \mathcal{L}_{Planck},
\label{Eq_LogLike_Total}
\ee
where the terms on the right hand side represent contributions from EDGES, ARCADE2 and {\it Planck}, respectively.  We discuss each term below.

\subsection{EDGES}
We adopt a Gaussian form for the EDGES likelihood,
\be
\ln \mathcal{L}_{\r{EDGES}}
=
-\frac{1}{2}
\sum
\left[
\frac{
(T_\r{sky}
-
T_\r{model})^2
}
{
\sigma^2_\r{T}}
+
2 \ln \sigma_\r{T}
\right]
+
const,
\label{Eq_LogLike_EDGES}
\ee
where the summation is over all weighed EDGES frequency bins,
and we treat the Gaussian error $\sigma_\r{T}$ as a free parameter to be varied in our inference.
$T_{\rm{sky}}$ is the EDGES sky temperature data discussed in \Sec{Sec_EDGES_Observation},
and the theoretical model value $T_{\rm{model}}$ is a combination of the 21cm signal $\bar{T}_{21}$,
foreground temperature $T_\r{FG}$ and residual calibration systematics $T_\r{cal}$,
\be
T_\r{model}
=
\bar{T}_{21}
+
T_\r{FG}
+
T_\r{cal}.
\label{eq:skytemp}
\ee

\begin{table*}[htp]
\centering
\begin{tabular}{c|c|ccccc}
\hline
Component & Parameters & \ \ Eq. & \ \ Units \ \ & \ \ Flat prior \ \ & \ \ Allowed range\\
\hline
$\bar{T}_{21}$ & 
$f_\r{R, III}$ & \ref{Eq_EMS_Radio} & - & log & [-2, 6]\\
 & $\mathcal{L}_\r{X, III}$ &\ref{Eq_LX} & $\r{erg\ s^{-1} M^{-1}_{\odot} yr}$ & log & [33, 45]\\
 & $f_\r{esc, 7}$ & \ref{Eq_Fesc_III} & - & log & [-6, -1]\\
 & $f_\r{\star, 7}$ &\ref{Eq_Fstar_III} & - & log & [-5, 0]\\
 & $A_\r{LW}$ & \ref{Eq_LW}  & - & linear & [0, 10]\\
\hline
$T_\r{FG}$ &
$p_0$ & \ref{Eq_TFG}  & K & linear & [$1000$, $3000$]\\
 & $p_1$ & \ref{Eq_TFG}  & K & linear & [-2000, 2000]\\
 & $p_2$ & \ref{Eq_TFG}  & K & linear & [-$10^3$, $10^3$]\\
 & $p_3$ & \ref{Eq_TFG}  & K & linear & [-$10^3$, $10^3$]\\
 & $p_4$ & \ref{Eq_TFG}  & K & linear & [-$10^3$, $10^3$]\\
 & $p_5$ & \ref{Eq_TFG}  & K & linear & [-$10^4$, $10^4$]\\
 & $p_6$ & \ref{Eq_TFG}  & K & linear & [-$10^4$, $10^4$]\\
 & $p_7$ & \ref{Eq_TFG}  & K & linear & [-$10^5$, $10^5$]\\
 & $p_8$ & \ref{Eq_TFG}  & K & linear & [-$10^6$, $10^6$]\\
 & $p_{9}$ & \ref{Eq_TFG}  & K & linear & [-$10^6$, $10^6$]\\
\hline
$T_\r{cal}$&
$a_{0}$ & \ref{Eq_Tcal}  & K & log & [-10, 2]\\
 & $a_{1}$ & \ref{Eq_Tcal}  & K & log & [-10, 2]\\
 & $P$ & \ref{Eq_Tcal}  & MHz & linear & [10, 15]\\
\hline
$\sigma_{\r{T}}$ &
$\sigma_{\r{T}}$ & \ref{Eq_LogLike_EDGES}  & K & log & [-4, -1]\\
\hline
\end{tabular}
\caption{
Parameters varied in our inferences and their allowed range,
we follow \citetalias{Sims:2019kro} to set priors for $\sigma_\r{T}$ and $T_\r{cal}$.
}
\label{Tab_varied_params}
\end{table*}

\subsection{ARCADE2}

Our ARCADE2 likelihood penalises models that produce radio excess above the observed level, which includes a contribution from galactic sources as well as a potential cosmological component.  
Since the measurement is therefore an upper limit on the cosmological component, we adopt a simple one-sided Gaussian form\footnote{
A more rigorous treatment for upper bounds in likelihood analysis can be found in 
\citealt{RuizdeAustri:2006iwb}.}:
\be
\begin{aligned}
\ln {\mathcal{L}}_\r{ARCADE2}
=&
-\frac{1}{2}
\sum
\frac{
(\bar{T}_{\r{radio}} - T_{\r{ARCADE2}})^2
}
{\sigma^2_\r{ARCADE2}}
\\&
\times
\Theta(\bar{T}_{\r{radio}} - T_{\r{ARCADE2}})
+
const,
\end{aligned}
\label{Eq_LogLike_ARCADE2}
\ee
where the summation is performed over the simulated redshifts ($3.8 <z < 36$),
$\Theta(x)$ is the Heaviside step function which equals unity for $x>0$ and vanishes otherwise,
$T_{\r{ARCADE2}}$ is the ARCADE2 excess level at 1.429GHz  and redshift $z$,
which can be derived from Eqs. (\ref{Eq_ARCADE_z0}) and (\ref{Eq_ARCADE_model_conatraints}) as,
\be
T_{\r{ARCADE2}}
=
1.19
(1+z)
\left(
\frac{v'_{21}}
{1 \r{GHz}}
\right)^{-2.62}
,\ 
v'_{21} = \frac{1.429 \ \r{GHz}}{1+z}.
\label{Eq_T_ARCDDE_v21}
\ee
Following the justification detailed in appendix \ref{Appdx_ARCADE2_Posterior},
we approximate the uncertainty as:
\be
\sigma_\r{ARCADE2}
=
0.1T_\r{ARCADE2}.
\ee

\subsection{{\it Planck}}
The reionisation optical depth $\tau_\r{rei}$ is constrained by CMB measurements from the  {\it Planck} 2018 measurements ~(see \citealt{Qin:2020xrg}) to be:
\be
\tau_\r{rei}
=
0.0569^{+0.0081}_{-0.0066}.
\label{Eq_tau_Limits}
\ee
We follow~\citealt{Greig:2018hja} to construct our {\it Planck} optical depth likelihood as,
\be
\ln {\mathcal{L}}_{Planck}
=
-\frac{1}{2}
\left[
\frac{
(\tau_\r{rei} - \tau_\r{plk})^2
}
{\sigma_u \sigma_l
+
(\sigma_u - \sigma_l)
(\tau_\r{rei} - \tau_\r{plk})
}
\right]
+
const,
\label{LogLike_tau}
\ee
where $\tau_\r{plk} = 0.0569$,
$\sigma_\r{u} = 0.0081$ and $\sigma_\r{l} = 0.0066$.

\section{Inference Methodology}
\label{Sec_Inference}

\subsection{Models}

In our inference we vary five free parameters characterizing Pop III galaxies:
\be
\{
f_\r{R, III}, 
\mathcal{L}_\r{X,III},
f_\r{esc, 7},
f_{\star, 7},
A_\r{LW}
\}.
\label{Eq_Varied_Params}
\ee
Our cosmological model and associated prior ranges are fixed, and we explore several models varying combinations of:
(i) polynomial order for FG parameters (from 4th-10th order);
and 
(ii) inclusion of sinusoidal calibration residuals (with/without residuals).
These combinations yield a total of 14 models,
which are then compared using the Bayesian evidence.  We summarize all free parameters and their prior ranges in \Tab{Tab_varied_params}.

\begin{figure*}[htp]
\centering
\includegraphics[width=18cm]{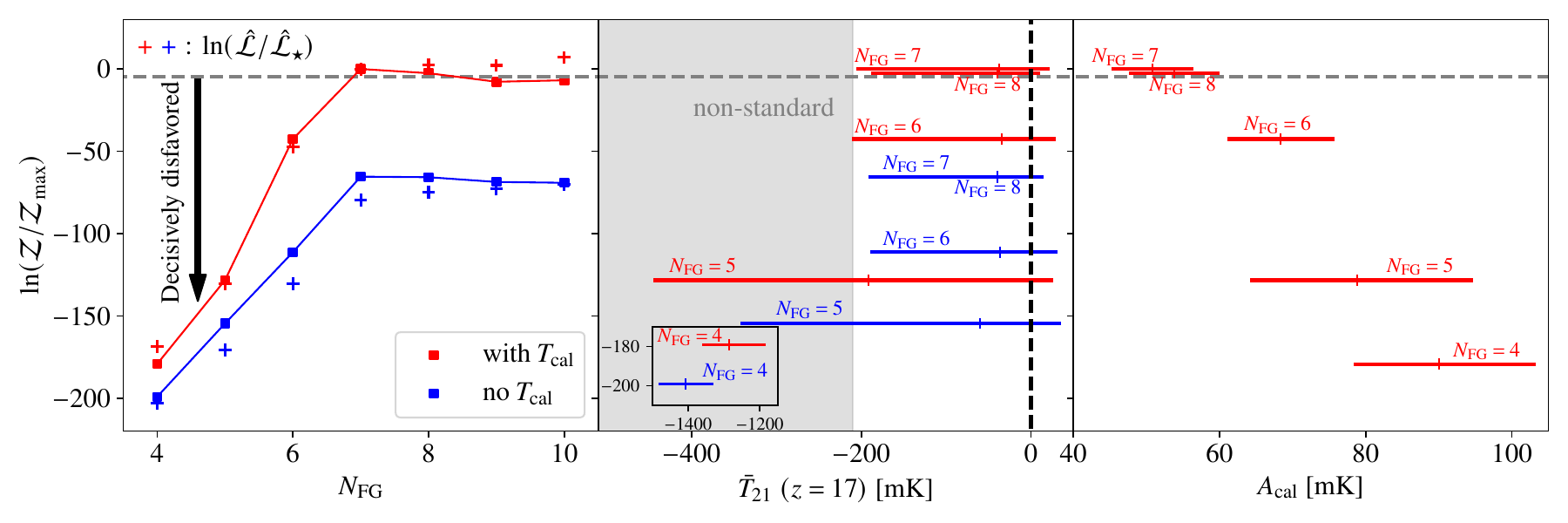}
\caption{
Bayesian log-evidence normalized to the  highest evidence model, $\ln \mathcal{Z/Z}_\r{max}$, as a function of foreground flexibility $N_\r{FG}$ ({\it squares in the left panel}),
$\Tbb$ at $z=17$ ({\it middle}) and calibration amplitude $A_\r{cal}$ ({\it right}, defined as $A_\r{cal} \equiv \sqrt{a_0^2 + a_1^2}$).
  Models with $N_\r{FG} \ge 9$ are omitted in the middle and right panel for visibility, as they overlap with $N_\r{FG} = $ 8, 9.
Models below the horizontal gray dashed line are decisively disfavored ($\ln \mathcal{Z/Z}_\r{max} \le -4.6$).
We use red and blue colors to represent scenarios of {\tt no $T_\r{cal}$} and {\tt with $T_\r{cal}$}, respectively.  The error bars in the middle and right panels represent 95\% C.I. regions around the mean.
In the middle panel,
the region below the standard expectation ($\bar{T}_{21} < -210$mK) is shaded in gray,
while the vertical black dashed line indicates $\Tbb=0$.
In the left panel,
the blue and red crosses show the maximum log-likelihood ratio, 
$\ln (\hat{\mathcal{L}}/\hat{\mathcal{L}}_{\star})$, 
normalized to our highest evidence model.
}
\label{Fig_Bayesian_Evidence}
\end{figure*}

Inference for each of these models varies 10-19 parameters and takes between 0.4M to 60M likelihood calls depending on the model complexity.
For computational convenience,
in all our inferences and relevant post-processing,
we calculate $\Tbb$, $\bar{T}_\r{radio}$, $\tau_\r{rei}$ and $\Delta^2_{21}$ using the \texttt{21cmEMU} emulator.  We train this emulator on a total of $1.3 \times 10^5$  {\tt 21cmFAST} outputs.  After training, the emulator error on the output is orders of magnitude smaller than the corresponding observational uncertainty, and is therefore negligible.
The details of the training process and the dataset are provided in Appendix \ref{Appdx_Emulator}.

\subsection{Model comparison using the Bayesian evidence}
For a model $\mathcal{M}$ characterized by a set of parameters $\theta$,
Bayes theorem states that given the data $D$,
the posterior probability distribution of $\theta$ is given by,
\be
P({\boldsymbol \theta} | D,\mathcal{M})
=
\frac{
\mathcal{L}
(D|\mathcal{M}, {\boldsymbol \theta})
\pi({\boldsymbol \theta}|\mathcal{M})}
{\mathcal{Z}
(D|\mathcal{M})},
\ee
where $\mathcal{L}$ and $\pi$ are likelihood and prior, respectively, 
and the Bayesian evidence $\mathcal{Z}$ is given by:
\be
\mathcal{Z}
(D|\mathcal{M})
=
\int 
\mathcal{L}
(D|\mathcal{M}, {\boldsymbol \theta})
\pi({\boldsymbol \theta})
\r{d} {\boldsymbol \theta}.
\ee

For parameter inference, the Bayesian evidence is simply a constant serving to normalize the posterior probability density, $P$, and can therefore be neglected.  However, in the presence of different models or prior distributions, the {\it relative} evidences quantify the degree by which our prior model odds are changed by the data~(see e.g.~\citealt{Trotta_2008}). Denoting by $\pi(\mathcal{M}_i)$ the prior probability for model $\mathcal{M}_i$, the posterior odds between two models $\mathcal{M}_i, \mathcal{M}_j$ is given by:
\be
\frac{P(\mathcal{M}_i | D)}{P(\mathcal{M}_j | D)} = \frac{Z_i}{Z_j}\frac{\pi(\mathcal{M}_i)}{\pi(\mathcal{M}_j)} \, .
\ee
The ratio of evidences is called the {\em Bayes factor}, $B_{ij} = Z_i/Z_j$, and it incorporates a quantitative notion of Occam's razor: a model is preferred (i.e., has a larger Bayesian evidence) when it explains the data better (i.e., it achieves a larger maximum likelihood) with an economy of free parameters. Therefore, a model that can better reproduce the data (having a higher likelihood) and is more predictive (having a smaller number of free parameters or more constrained prior ranges compared to the posterior) results in a correspondingly higher Bayesian evidence.
When comparing between models $\mathcal{M}_1$ and $\mathcal{M}_2$,
we follow~\citealt{jeffreys1961theory,re1995bayes} and use $\ln \mathcal{Z}_1/\mathcal{Z}_2 > 4.6$ as `decisive' preference of $\mathcal{M}_1$ over $\mathcal{M}_2$.  In our analysis below, we assume a flat prior over all models; therefore the posterior odds between two models reduces to the corresponding Bayes factor.

\subsection{Nested sampling with {\tt MultiNest}}

Inferences in this work are performed using the {\tt MultiNest} ~\citep{Feroz:2008xx,Buchner:2014nha} package.
{\tt MultiNest} uses nested sampling ~\citep{Skilling:2004pqw} to efficiently calculate the Bayesian evidence and parameter posteriors.
We use default parameter choices, except for the number of live points ($N_\r{live}$) 
, which we set to $100n_\r{dim}$ in our fiducial inferences,
where $n_\r{dim}$ is the dimensionality of model parameter space.
We analyze the {\tt MultiNest} output using the {\tt GetDist}~\citep{Lewis:2019xzd} package to derive posteriors for model parameters and associated observables (e.g. $\Tbb$, $\bar{T}_\r{radio}$),
which are treated as derived parameters of our model.
We confirm that the posteriors have converged for the above choices.

\section{Results}
\label{Sec_Results}

\begin{figure*}[htp]
\centering
\includegraphics[width=18cm]{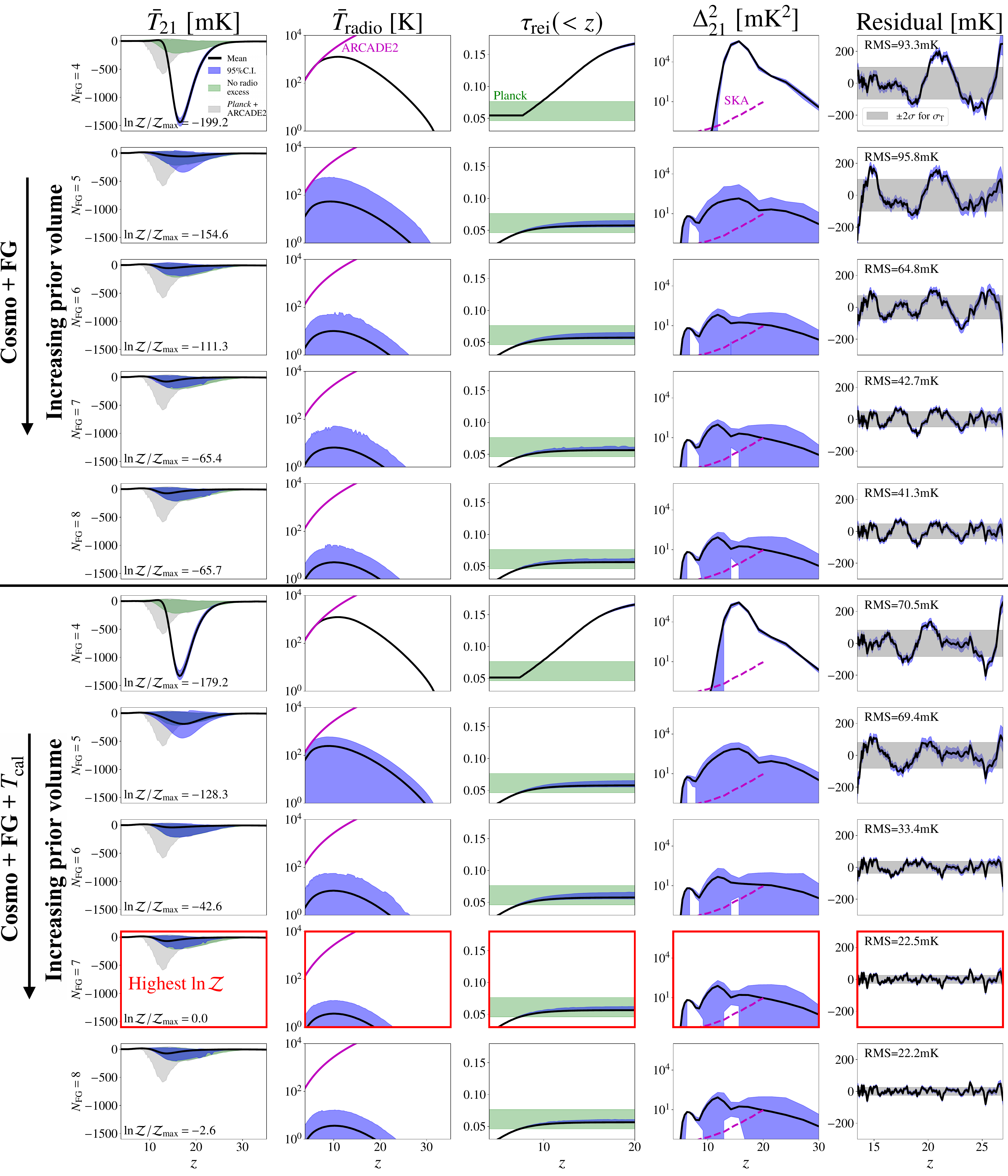}
\caption{
Marginalised posteriors from a subset of our inferences.
From left to right,
we show results for $\Tbb$,
$\bar{T}_\r{radio}$,
cumulative optical depth $\tau_\r{rei}(z)$,
$\Delta^{2}_{21}$ at $k = 0.1 \r{Mpc^{-1}}$,
residual ($T_\r{sky} - \Tbb - T_\r{FG} - T_\r{cal}$), respectively.
Each row corresponds to a different model, with the general trend of increasing systematics prior volume with decreasing row.  The number of foreground terms $N_\r{FG}$ is indicated in the y-axis in the $\Tbb$ sub-panels.
Black solid lines and blue shaded contours show mean and 95\% C.I. regions, respectively.
In the $\Tbb$ sub-panels,
the green contours indicate the 95\% C.I. assuming no radio background excess (i.e. setting $f_\r{R, III} = 0$ and only using the {\it Planck} likelihood), while 
the gray contours indicate the 95\% C.I. using only 
ARCADE2+{\it Planck} likelihoods (i.e. no EDGES).
The green bar in the $\tau_\r{rei}(z)$ sub-panels denotes the 95\% C.I. from {\it Planck}~\citep{Qin:2020xrg} (see \Eq{Eq_tau_Limits}), while
 the magenta dashed lines in the $\Delta^2_{21}$ sub-panels shows projected SKA noise power at $k=0.1\r{Mpc}^{-1}$ with a 1000 hour integration taken from ~\citealt{Barry:2021szi}.
In the residual sub-panels, the $\pm 2\sigma$ extent of the inferred Gaussian error $\sigma_\r{T}$ is demarcated in gray.}
\label{Fig_main_posteriors}
\end{figure*}

\begin{figure}[htp]
\centering
\includegraphics[width=8.5cm]{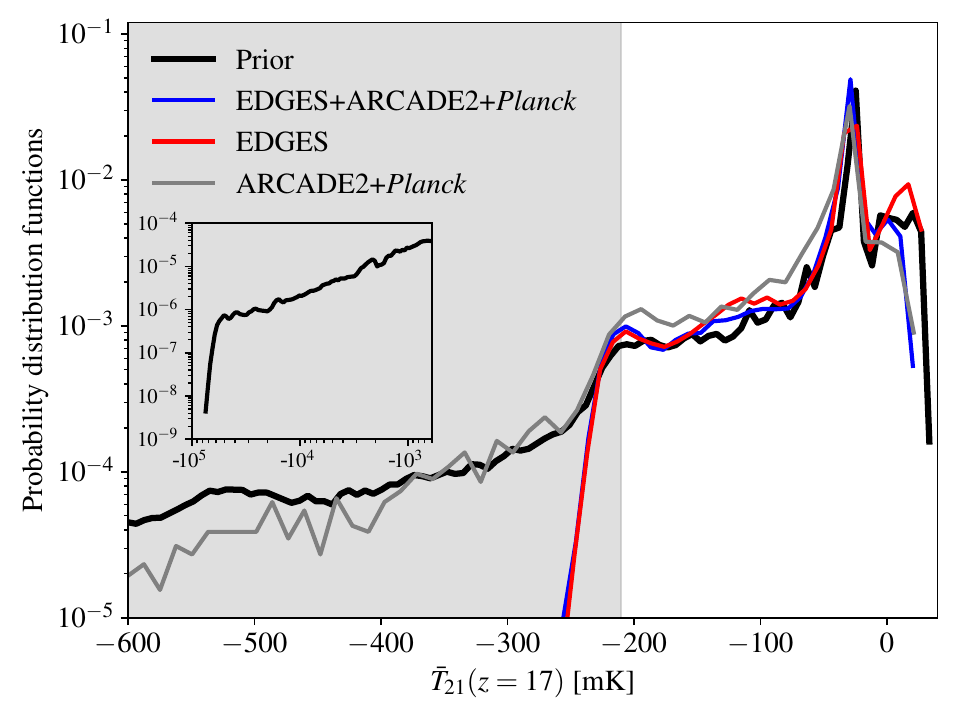}
\caption{
 PDFs for $\Tbb$ at $z=17$ from our highest evidence model.  The black curve corresponds to our prior distribution, while the other curves correspond to posterior PDFs obtained using various likelihood terms as denoted in the legend.
The gray shaded region marks the regime of non-standard depths.  The inset shows the prior extended to very low temperatures.}
\label{Fig_T17_Plot}
\end{figure}

We now show our main results,
in which we forward model the full EDGES sky temperature data, constructing the likelihood directly in data space as discussed in \Sec{Sec_Inference}.
In \Fig{Fig_Bayesian_Evidence} we show the Bayesian evidence ratios normalized to the highest evidence model with $\mathcal{Z} = \mathcal{Z}_\r{max}$,
which features 7 FG terms and calibration residuals.
From {\it left} to {\it right},
in each panel we show results respectively as a function of (i) the number of FG terms,
(ii) $\Tbb$ at redshift 17, and (iii) the calibration amplitude $A_\r{cal}$ (defined as $\sqrt{a_0^2 + a_1^2}$).
In the middle panel, the gray shaded region indicates the regime of $\Tbb < -210$~mK, corresponding to models with a radio background in excess of the CMB.
For simplicity, 
hereafter we will refer to regimes with and without radio excess as non-standard and standard, respectively.

As can be seen from the left panel,
the Bayesian evidence peaks at $N_\r{FG}=7$ for both calibration settings. 
On the same plot, we also display the maximum log-likelihood value for each model, 
normalized to the maximum log-likelihood of the 
highest evidence 
model (blue and red crosses). 
We observe that the maximum log-likelihood increases monotonically with the number of free parameters, as expected, but that the slope decreases sharply at $N_\r{FG} = 7$. This indicates that the sharp increase in log-evidence until $N_\r{FG}=7$ is a reflection of the increasing quality of fit of models with more foreground free parameters. After $N_\r{FG}=7$, however, adding extra freedom to the model results in an Occam's razor penalty in the evidence, which manifests as a decrease in its value.  
In fact the only model that is not decisively disfavored in comparison to the highest evidence model is one with $N_\r{FG}=8$ and calibration error.
Thus calibration errors are needed to explain the data in the context of our physical model.

The middle panel of \Fig{Fig_Bayesian_Evidence} shows that at $z=17$ (i.e. the central redshift of the putative cosmic signal recovered by \citetalias{Bowman:2018yin}),
only  models with $N_\r{FG} \le 5$ have a preference for a radio background excess.  Adding more foreground terms {\it disfavors} non-standard $\Tbb$ amplitudes and {\it increases} the Bayesian evidence.  
For example, 
at $N_\r{FG}=4$,
$\Tbb$ amplitudes at $z=17$ are constrained to [-1482, -1183]~mK,
with corresponding log-evidence ratios between $-199$ and $-179$.  In contrast,
posteriors for all $N_\r{FG} \ge 6$  models lie exclusively in the standard range of $[-210, 30]$~mK.

In the right panel of \Fig{Fig_Bayesian_Evidence},
we find that inflexible FG models ($N_\r{FG} \leq5$) paired with our 21cm model are wildly insufficient to account for the data, 
and the calibration residuals are required to be very large ($A_\r{cal} \sim$103 mK). 
As the number of FG terms are increased,
some of the higher-order residual structure that was fit by the calibration-residual term is accounted for by the FGs, 
and the calibration-residual amplitude required reduces to $A_\r{cal}\sim 51 \r{mK}$ for the high-evidence models. 

In \Fig{Fig_main_posteriors} we show additional observables from these models.  The columns correspond to $\Tbb$, $\bar{T}_\r{radio}$,
$\tau_\r{rei}(<z) \equiv \int_0^{z} \frac{\rd \tau_\r{rei}} {\rd z'}\rd z'$,
$\Delta^2_{21}$, and the associated residuals from the EDGES observation, from left to right.
Each row corresponds to a different model, following the general trend of increasing prior volume with decreasing row.  The top subsection includes models without calibration residuals, while models in the bottom subsection include calibration residuals.  In both we show models with $N_\r{FG}$ = 4 -- 8.
In each subpanel, the mean and 95\% C.I. are indicated with black solid curves and blue shaded contours.  The green shaded contours in the $\Tbb$ subpanels correspond to the 95\% C.I. of "standard models" without a radio excess (i.e. obtained by setting $f_\r{R, III}$ to zero and using only the {\it Planck}  likelihood term).  The gray shaded regions in the same panels correspond to the 95\% C.I. using only the ARCADE2 and {\it Planck} likelihood terms (i.e. without EDGES).
In the $\bar{T}_\r{radio}$ and $\tau_\r{rei}$ subpanels,
we also show the ARCADE2 radio excess (see \Eq{Eq_T_ARCDDE_v21}) and the {\it Planck} 95\% C.I. constraint (derived from \Eq{LogLike_tau}) with magenta lines and green shaded regions, respectively.
The magenta dashed lines in the $\Delta^2_{21}$ panels show the thermal noise power at $k = 0.1 \r{Mpc}^{-1}$ expected from a 1000h integration with SKA  \citep{Barry:2021szi}.

As can be seen from the right-most column, the amplitude and the level of coherent (not noise-like) structure of the residuals decrease as we increase the prior volume of the systematics terms. {\it All} models that do not include calibration errors ({\it first five rows}) have visible structure in the residuals.  In order not to have structure in the residuals, we require $N_\r{FG} \geq 7$ AND calibration errors ({\it bottom two rows}).  
This again highlights that our physical model for the extra radio background, contrary to the ad-hoc flattened Gaussian, is unable to explain the EDGES signal by itself.

As was already noted in Fig. \ref{Fig_Bayesian_Evidence}, the only models that prefer a non-standard depth for $\Tbb$ are those with $N_\r{FG} \leq 5$.  Interestingly, these models require a very early reionization. This is because reionization can provide a more rapid rise in the global signal at $z\lesssim18$ compared with X-ray heating.  The improved agreement with the EDGES data from such a rapid rise compensates for its worse agreement with {\it Planck}.  Regardless, these models do not actually explain the EDGES data, as can be seen visually from the structure in the residuals and as is quantified by  the Bayesian evidence.
 
In red we highlight our highest evidence model, which includes seven foreground terms as well as calibration errors.  We see that the radio background and the CMB optical depth of this model are consistent with observations.  Moreover, the global signal residuals in the rightmost panel are noise-like.  Interestingly though, the posterior of the global signal seems perfectly aligned with the "no radio excess" posterior in green.  Indeed, this claim holds true for all models with $N_\r{FG} \geq 6$.  This implies that higher order foreground terms (and systematics errors) do a better job of reproducing the EDGES signal than our physical, excess radio background model.

We investigate this claim further in \Fig{Fig_T17_Plot}, showing the probability distribution function (PDF) of $\Tbb$ at redshift 17, for our maximum evidence model (a corner plot for this model is shown in Appendix \ref{Appdx_CornerPlot}).
Non-standard depths are demarcated with the gray shaded region.
The black line corresponds to our astrophysical prior, and is generated using $3.2\times10^6$ samples.
The other lines show posterior PDFs using different likelihood combinations, as indicated in the figure legend.  For these, we increase $N_\r{live} = 500 n_\r{dim}$, confirming that the posterior in the tails of the distribution has converged.

Roughly 17\% of our prior samples have non-standard depths ({\it black curve}). 
There is not much difference between the black and gray curves, showing that ARCADE2 and {\it Planck} do not have much impact on our prior in this space.\footnote{The slight drop of the PDF in gray compared to the prior in black at the left edge of the figure is due to the fact that these extreme models can also include extremely efficient star formation and/or high radio luminosities.  Such extremes could result in an early EoR and/or a radio excess that can already be disfavored by {\it Planck} and ARCADE2.}
However, when we include also the EDGES likelihood ({\it blue curve}), the posterior PDF shifts to {\it disfavor} non-standard models.
Roughly the same result is obtained if we only consider the EDGES data ({\it red curve}).
This figure shows that, not only does our radio excess model fail in explaining EDGES, it actually interferes with the systematics terms that do a better job.  In other words, the EDGES data {\it disfavors} a non-standard radio background during the Cosmic Dawn.  This is {\it qualitatively contrary} to previous conclusions, highlighting the inherent dangers of using a pseudo-likelihood built on a model (flattened Gaussian) that is different from the one being used in the inference \citep{mirocha2019does,Fialkov:2019vnb,Reis:2020arr,Ewall-Wice:2019may}.

\section{Discussion and caveats}
\label{Sec_Discussion_caveats}

Other works have also performed Bayesian analysis on EDGES data in order to further test the \citetalias{Bowman:2018yin} result (e.g. ~\citealt{Hills:2018vyr,Singh:2019gsv,Sims:2019kro,Tauscher:2020wso,Murray:2022uhg}).
Most however either perform inference with a flattened Gaussian, or use the flattened Gaussian to define a pseudo-likelihood~\citep{mirocha2019does,Fialkov:2019vnb,Ewall-Wice:2019may,Reis:2020arr}.
Closest to our analysis was \citetalias{Sims:2019kro} in which the authors compared Bayesian evidences of multiple models for the cosmic signal, FGs, calibration residuals and Gaussian noise ($\sigma_\r{T}$ of our EDGES likelihood in \Eq{Eq_LogLike_EDGES}).
They concluded that the EDGES signal could be explained either with a phenomenological flattened Gaussian with a non-standard depth or with calibration residuals.
Their physical models for the cosmic signal (computed with the ARES code;~\citealt{Mirocha:2014faa}) did not include a radio background excess.  Here we show that a physical model for the radio background excess, as opposed to a phenomenological flattened Gaussian, is actually disfavored by EDGES data.

While we have shown for our specific cosmological model that EDGES intrinsically disfavors a non-standard $\Tbb$ depth, we expect a similar result for {\it any} physical model anchored on galaxy evolution.
Though EDGES data has been shown to give non-standard depth for the phenomenological flat Gaussian $\Tbb$ model (see e.g. \citetalias{Bowman:2018yin} and the updated analysis in \citealt{Murray:2022uhg}),
there are two unusual features of such a profile that cannot be simultaneously mimicked by realistic astrophysical models: the flatness of the trough and the sharpness of the wings.

All realistic cosmological explanations of the EDGES signal require galaxies and/or black holes to play a role during the cosmic dawn (e.g.~\citealt{mirocha2019does,Reis:2020arr,Ewall-Wice:2019may}).
The redshift evolution of galaxies is linked to the well-known evolution of the halo mass function,
which cannot mimic the sharp features of the flattened Gaussian shape reported in \citetalias{Bowman:2018yin} while maintaining the consistency with other astrophysical constraints.
\citealt{Kaurov:2018kez} suggested that the sharpness can be reproduced if Lyman-alpha coupling between spin and kinetic temperatures were driven by halos with masses in excess of $\sim 10^{9-10}M_\odot$, and assuming a constant mass to light ratio.
However, there is no known physical mechanism to quench star formation in smaller halos at $z>18$; moreover, we know from observed UV LFs that galaxies do not have constant mass to light ratios (see e.g.~\citealt{Bromm:2013iya,Park:2018ljd}).  Even with these caveats, such a model for a steep evolution of the global signal does not result in a flattened shape.

The flatness in the flattened Gaussian shape could be achieved if there are multiple IGM heating sources that are very precisely tuned so that their combination heats the gas temperature as $T_\r{K} \propto (1+z)$.  IGM heating that is comparable to that provided by X-rays could be achieved in some models of dark matter decay (e.g.~\citealt{Facchinetti:2023slb,Sun:2023acy}), annihilation (e.g.~\citealt{Valdes:2012zv,Evoli:2014pva,Lopez-Honorez:2016sur,Cang:2023bnl}) and cosmic rays (e.g. ~\citealt{Leite:2017ekt,Gessey-Jones:2023amq}).
Such a multiple heating scenario was in fact suggested by~\citealt{Gessey-Jones:2023amq},
in which the authors implemented both X-ray and cosmic ray heating (see their Fig.10).
However this requires extreme fine-tuning of astrophysics in order to achieve such a precise evolution, {\it in addition} to the already exotic assumption of a radio background excess; thus the Bayesian evidences of such models would be strongly penalized by the small prior volume of viable parameters.  Moreover, even with a relatively flat evolution resulting from such astrophysical "tuning", a double heating scenario is unable to achieve the sharpness of the flattened Gaussian MAP from \citetalias{Bowman:2018yin}, as also pointed out in \citealt{Gessey-Jones:2023amq}.

Finally we point out that our results also depend on our specific model and associated prior volume for foregrounds and systematics.   As mentioned earlier, both of these are physically-motivated.  However, our understanding of systematics is far from complete.  With an improved characterization of FG+calibration residuals resulting in a more physical basis and/or prior volume, our conclusions could change.

\section{Conclusions}
\label{Sec_Conclusions}

The existence of a radio background at $z\sim17$ in excess of the CMB provided a popular explanation for the strong global 21cm  signal reported by the EDGES collaboration (e.g. \citealt{Feng:2018rje,ewall2018modeling,Sharma:2018agu,Fialkov:2019vnb,mirocha2019does,Ewall-Wice:2019may,Reis:2020arr,Ziparo:2022fnc,Sikder:2022hzk,Zhang:2023rnp,Sikder:2024osn}).
This work investigates the radio excess from the first generation Pop III galaxies.
These galaxies are naturally sterilized at lower redshifts due to LW and photoheating feedback,
and thus can reproduce the EDGES absorption depth without violating upper limits on the radio background from ARCADE2.

We demonstrate that Pop III radio galaxies can indeed drive a 21cm absorption signal that has the same depth and timing of the phenomenological flattened Gaussian recovered by EDGES, without violating constraints from complimentary observations. 
We perform Bayesian inference on EDGES sky temperature data, together with constraints from ARCADE2 and {\it Planck}.  Our models for the sky temperature consist of the cosmic signal combined with foreground and calibration errors of varying complexity.  We compare these models using their Bayesian evidence.

Models that do not account for calibration errors are decisively disfavored by the data, showing clear structure in the residuals (i.e. the difference between the observed and forward-modeled sky temperature).
Our highest evidence model is characterized by seven log-polynomial foreground terms and calibration residuals. All models that have "noise-like" residuals and are not decisively disfavored by the EDGES data have posteriors consistent with standard model predictions (i.e. without a radio background excess). We show that not only does our radio excess model fail in explaining EDGES, but that excess radio backgrounds that produce beyond-standard absorption depths actually interfere with systematics models that do a better job.

Our conclusion that EDGES {\it disfavors} a strong cosmic 21cm signal is different from all previous works that simulated a radio background excess.  This difference stems from the fact that here we forward model the EDGES temperature data directly and use a physical model for the cosmic 21cm signal.   Previous analyses that included physical models for the excess used a "pseudo-likelihood" based on a different model (the phenomenological flattened Gaussian).  Our work therefore serves to highlight the importance of self-consistent inference.  We make our simulation code\footnote{
\href{https://github.com/Junsong-Cang/21cmFAST/tree/Radio_Excess}{https://github.com/Junsong-Cang/21cmFAST/tree/Radio\_Excess}}
and associated emulator\footnote{
\href{https://github.com/21cmfast/21cmEMU}{https://github.com/21cmfast/21cmEMU}} publicly available.

\begin{acknowledgements}
The authors gratefully acknowledge the HPC RIVR consortium and EuroHPC JU for funding this research by providing computing resources of the HPC system Vega at the Institute of Information Science (project EHPC-REG-2022R02-213). We thank Peter Sims, James Davis and Ivan Nikolić for helpful discussions.
J.C. acknowledges support by SNS and the China Scholarship Council.
A.M. and R.T. acknowledge support from the Ministry of Universities and Research (MUR) through the PRIN project "Optimal inference from radio images of the epoch of reionization", the PNRR project "Centro Nazionale di Ricerca in High Performance Computing, Big Data e Quantum Computing", and the PRO3 Scuole Programme `DS4ASTRO'.
This project has received funding from the European Union’s Horizon 2020
research and innovation programme under the Marie Skłodowska-Curie grant
agreement No 101067043. Y.Q. is supported by the ARC Discovery Early Career Researcher Award (DECRA) through fellowship \#DE240101129.
\end{acknowledgements}

\bibliographystyle{aa}
\bibliography{main.bib}

\appendix
\section{Using a "pseudo-likelihood" based on a backward-modeled flattened Gaussian}
\label{Sec_Pseudo_Likelihood}

\begin{figure*}[ht]
\centering
\includegraphics[width=18cm]{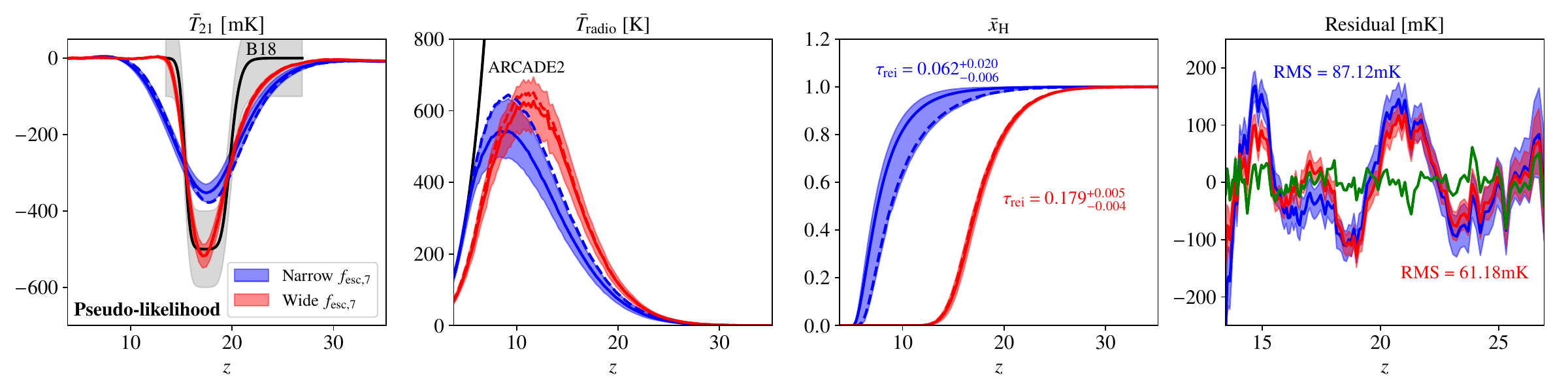}
\caption{
Results from our pseudo-likelihood inferences,
in which the likelihood is defined on the recovered flattened Gaussian from  ~\citetalias{Bowman:2018yin}, instead of directly in the sky temperature space (see text for details).
Our ARCADE2 and {\it Planck} likelihoods are also included in the analysis.
From left to right,
we show in red and blue the posteriors for $\Tbb$, $\bar{T}_\r{radio}$, $\bar{x}_\r{H}$ and the residuals.
Solid and dashed lines correspond to mean and MAP, respectively,
and shaded contours indicate 95\% C.I.s.
Results in red use the astrophysical prior ranges listed in \Tab{Tab_varied_params} and lead to an extremely early reionization.  Those in blue adopt a narrow $f_\r{esc,7}$ prior of $[10^{-6}, 10^{-2.5}]$ to accommodate EoR timing constraints.
In the $\Tbb$ panel,
we show \citetalias{Bowman:2018yin} profile with black solid line and gray shaded region represents the uncertainty level adopted in our Pseudo-likelihood.
The black solid line in the $\bar{T}_\r{radio}$ panel shows the ARCADE2 excess.
The 95\% C.I. on $\tau_{\r{rei}}$ are indicated in the $\bar{x}_\r{H}$ panel.
For the residuals panel,
we fit the EDGES data with a five term FG model while fixing the cosmic 21cm signal to the corresponding MAP results in $\Tbb$ panel.  There is clear structure in the residuals.  For comparison, we also include the residuals from our highest evidence model as a green line, which are noise-like.}
\label{Fig_Likelihood_tests}
\end{figure*}

\begin{figure}[htp]
\centering
\includegraphics[width=9cm]{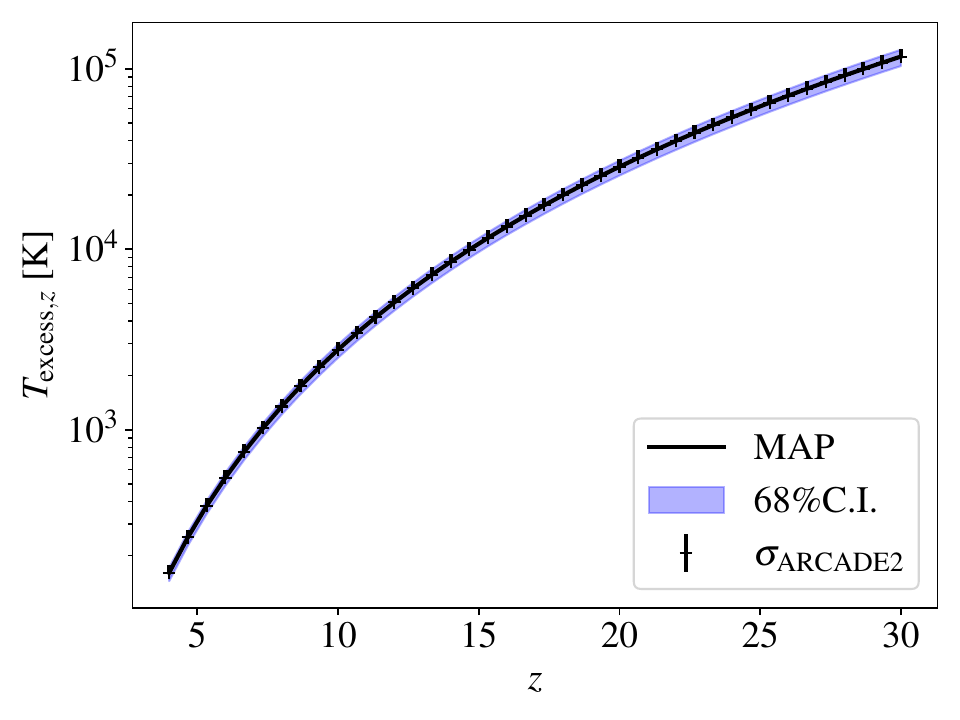}
\caption{
Posteriors of radio excess level inferred from ARCADE2 datasets,
the black solid curve shows the MAP results,
and blue filled contour shows the $68 \%$ C.I. region,
the black error bars show the error $\sigma_\r{ARCADE2}$ used for out ARCADE2 likelihood in \Eq{Eq_LogLike_ARCADE2}.}
\label{Fig_ARCADE2_Posteriors}
\end{figure}

\begin{figure*}[htp]
\centering
\includegraphics[width=18cm]{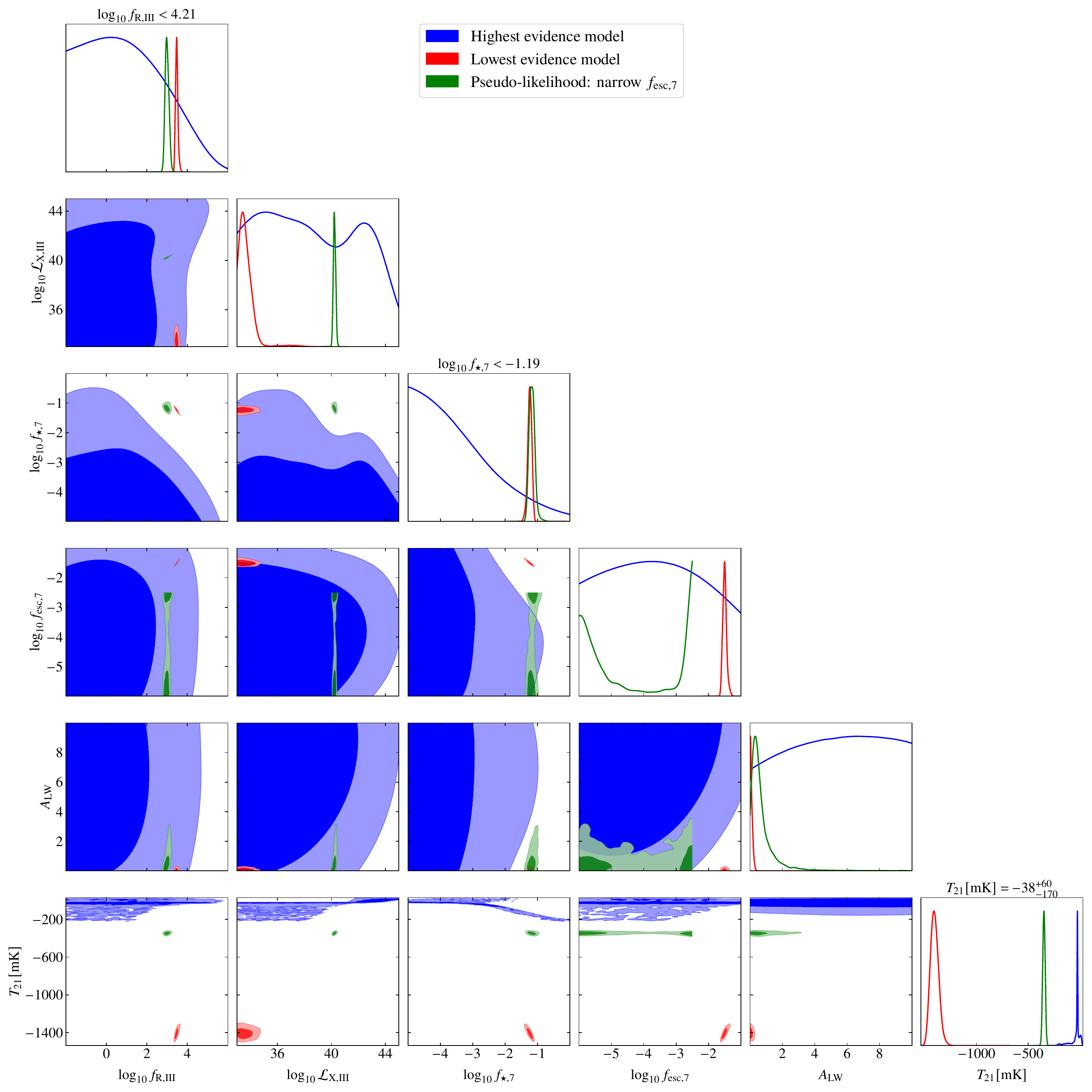}
\caption{
1D and 2D marginal posteriors for our astrophysical parameters and derived $\Tbb$ at $z=17$ resulting from different inferences.  
The plotted  parameter ranges corresponds to their respective prior ranges.
Blue and red contours and curves show results for our highest evidence model (seven FG terms, with calibration residuals) and lowest evidence model (four FG terms, no calibration residuals), respectively.
Posteriors for the {\tt Narrow} $f_\r{esc,7}$ pseudo likelihood inference from Appendix A are shown in green.
The headings above the panels show 95\%C.I. for the highest evidence model.
}
\label{Fig_Corner_Plot}
\end{figure*}

Here we show posteriors for a simplified version of the EDGES likelihood.
Instead of self-consistently forward modeling the EDGES sky temperature,
here we we construct the likelihood on the recovered (MAP) flattened Gaussian from \citetalias{Bowman:2018yin}.
This serves to mimic the common approach in the literature (e.g.~\citealt{mirocha2019does,Fialkov:2019vnb,Ewall-Wice:2019may,Reis:2020arr}), 
in which a physical cosmological model is compared with a recovered empirical profile (flattened Gaussian),
without directly forward-modeling the observed data.

Specifically, 
we follow \citealt{mirocha2019does} and construct a Gaussian pseudo-likelihood of the form,
\be
\ln \mathcal{L}
=
-
\sum_{\r{i}}
\frac{(T_{\r{B18, i}} - \bar{T}_{21,\r{i}})^2}
{2\sigma^2}
+
const,
\ee
where the summation is over all lightcone redshifts within EDGES frequencies,
$T_{\r{B}18}$ is the flattened-Gaussian profile reported in \citetalias{Bowman:2018yin},
we assume that different redshifts are independent and set the uncertainty $\sigma$ to 100mK.

In the first three panels of \Fig{Fig_Likelihood_tests} we show the posteriors for $\Tbb$, $\bar{T}_\r{radio}$ and $\bar{x}_\r{H}$ using this pseudo-likelihood.
We adopt the priors in \Tab{Tab_varied_params} for results shown in red (hereafter dubbed {\tt Wide $f_\r{esc,7}$}). 
From the $\Tbb$ and $\bar{T}_\r{radio}$ panels we see that
our model can roughly recover the \citetalias{Bowman:2018yin} result
without invoking an ad-hoc "$z_{\rm off}$" parameter to turn off the radio and/or X-rays from the first galaxies~\citep{mirocha2019does, Reis:2020arr,Sikder:2022hzk,Sikder:2024osn}.
Self-consistently modeling the feedback on Pop III galaxies is able to physically turn off this population so that their radio background does not exceed the ARCADE2 observations.

However,
the recovered optical depth for the above inference is significantly higher than inferred from {\it Planck} data.
This is due to the fact reionization can provide a more rapid rise in the global signal at $z\lesssim18$ compared with X-ray heating. 
The improved agreement with the flattened Gaussian pseudo-likelihood is enough to exceed the corresponding penalty from the {\it Planck} likelihood.
To accommodate {\it Planck} optical depth constraints (see \Eq{Eq_tau_Limits}),
we also perform inference restricting the prior range on the escape fraction $f_\r{esc,7}$.
Specifically, 
posterior in blue corresponds to a tighter log-uniform $f_\r{esc,7}$ prior of $[10^{-6}, 10^{-2.5}]$ (dubbed as {\tt Narrow $f_\r{esc,7}$} hereafter).
The recovered $\Tbb$ posterior of the {\tt Narrow $f_\r{esc,7}$} model is a lot less narrow compared to the {\tt Wide $f_\r{esc,7}$}  model.  Without reionization at $z\sim16$, the rise in the global 21cm signal is limited by the (comparably less efficient) EoH and is therefore slower.


The last panel of \Fig{Fig_Likelihood_tests} shows the residuals from fitting the EDGES data with only a 5 term FG while fixing $\Tbb$ to the MAP results from our pseudo likelihood fits.
The red and blue contours indicate 95\% C.I. regions for model using $\Tbb$ from {\tt Wide $f_\r{esc,7}$} and {\tt Narrow $f_\r{esc,7}$} set-ups, respectively.
If the combination of fixed cosmic signal and FG did describe the EDGES data,
one would expect the residuals to be noise-like.
However it can be seen that the residuals show oscillating structures that are not consistent with noise,
especially when compared with that of our highest evidence model (green dashed line, detailed in \Sec{Sec_Results}).
This indicates that the combination of FG and fixed \citetalias{Bowman:2018yin}-like cosmic signal (inferred from the pseudo-likelihood) does not give a reliable description to the actual EDGES data.
We therefore conclude that for physical models,
one cannot use a likelihood based on summaries of the recovered flattened Gaussian.

\section{Emulating the cosmic signal with \texttt{21cmEMU}}
\label{Appdx_Emulator}

The database used in our training consists of $1.3 \times 10^5$ {\tt 21cmFAST} simulations, distributed within our prior regions of astrophysical parameters (see \Tab{Tab_varied_params}).
Samples are generated with our customized version of {\tt 21cmFAST} and span a redshift range of $z \in [3.8, 36]$,
with a box of size 500 Mpc and a resolution of $50^3$ (which we find is sufficient for convergence in all of our observables to better than $\sim$ few percent).
This database is then split into training, validation and test sets,
each containing $10^5$, $10^4$ and $10^4$ samples, respectively.
The network has a tree-like architecture where the input is passed into separate branches that each output the prediction for one of $\Tbb$, $\bar{T}_\r{radio}$, $\bar{x}_\r{H}$, $\tau_\r{rei}$ and $\Delta^2_{21}$.
In \Tab{tab:emulator}, 
we illustrate the number of layers and the number of neurons per layer for $\Tbb$, $\bar{T}_\r{radio}$ and $\tau_\r{rei}$ branches,
the results of which enter directly into our likelihoods.

We implement the emulator using \texttt{PyTorch} \citep{Paszke19}. After training the emulator, we evaluate it against a test set. To measure its performance, we calculate the fractional error (FE):
\begin{align}
    &\rm{Abs~Diff} \equiv | y_{\rm true} - y_{\rm pred}|, \\
    &\rm{FE (\%)} \equiv \frac{ \rm{Abs~Diff} }{\max{(|y_{\rm true}|, y_{\rm floor}})} ,
\label{eq:FE}
\end{align}
where $y_{\rm true}$ refers to the {\tt 21cmFAST} simulation output and $y_{\rm pred}$ is the corresponding emulator prediction.
In the last two rows of Table \ref{tab:emulator}, we show the mean FE as well as its 68\% C.I. limit for each summary. We see that the brightness and radio temperatures have a mean FE of about 4\%, while the neutral fraction and reionisation optical depth have a mean FE of less than one percent.  These emulator errors are orders of magnitude smaller than the observational uncertainty, and are thus negligible.

\begin{table}[htp]
\centering
\begin{tabular}{c|ccc}
\hline
          & $\Tbb$   & $\bar{T}_\r{radio}$   & $\tau_\r{rei}$ \\ 
\hline
Number of layers  & 10   & 10   & 5    \\
Number of neurons & 1500 & 1000 & 500  \\
Mean FE (\%)   & 3.8  & 3.7  & 0.1     \\
FE 68\% C.I. (\%)   & 13.0 & 8.3  & 18.9\\
\hline
\end{tabular}
\caption{Parameters and performance for each branch of the emulator. The first two rows indicate the number of layers and neurons in each branch of the network. The last two rows indicate the mean fractional error and 68\% C.I. obtained from evaluating the emulator against the test set.}
\label{tab:emulator}
\end{table}

\section{Uncertainty level for ARCADE2 radio excess temperature}
\label{Appdx_ARCADE2_Posterior}

Our choice for the uncertainty level $\sigma_\r{ARCADE2}$ in \Eq{Eq_LogLike_ARCADE2} is well motivated by Bayesian analysis.
The current radio brightness temperature measurements reported by ARCADE2~\citep{Fixsen:2009xn} can be expressed as the sum of temperatures of CMB $T_\r{CMB}$ and radio excess $T_\r{excess}$,
\be
T = T_\r{CMB} + T_\r{excess},
\label{Eq_T_ARCADE2_Model}
\ee
and $T_\r{excess}$ can be parameterized by a power-law,
\be
T_\r{excess}
=
T_\r{r}
\left(
\frac{\nu}
{\r{GHz}}
\right)^{\beta}.
\label{Eq_T_ARCADE2_Excess_Model}
\ee

We treat $T_\r{CMB}$, $T_\r{r}$ and $\beta$ as free model parameters and derive their posterior distributions by sampling the following likelihood using {\tt MultiNest}~\citep{Feroz:2008xx,Buchner:2014nha},
\be
\ln \mathcal{L}
=
-\frac{1}{2}
\sum_i
\frac{T_\r{i} - T_\r{data,i}}
{\sigma_\r{i}^2}
+const
\ee
where the subscript i indicates frequency,
$T_\r{i}$ is given in \Eq{Eq_T_ARCADE2_Model},
the data $T_\r{data,i}$ and uncertainty $\sigma_\r{i}$ are taken from Table 4 of \citealt{Fixsen:2009xn}.

Our inference constrains the model parameters to $T_\r{CMB} = 2.730 \pm 0.004 \r{K}$,
$T_\r{r} = 1.198 \pm 0.129\r{K}$ and $\beta = -2.613 \pm 0.042$ (uncertainties represents 68\% C.I.),
which is in almost identical agreement with values quoted in \citealt{Fixsen:2009xn}.

Note that the excess level in \Eq{Eq_T_ARCADE2_Excess_Model} was given for $z=0$.
At 21cm frequency and arbitrary redshift $z$,
the relevant excess level can be expressed as,
\be
T_{\r{excess},z}
=
(1+z)
T_\r{excess}(\nu'_{21}),
\nu'_{21} = \frac{1.429\r{GHz}}{1+z},
\ee
which is determined by model parameters $T_\r{r}$ and $\beta$ following \Eq{Eq_T_ARCADE2_Excess_Model}.
The posterior distributions for $T_{\r{excess},z}$ can be easily derived by analyzing the {\tt MultiNest} inference chains for $T_\r{r}$ and $\beta$.

In \Fig{Fig_ARCADE2_Posteriors},
we show the MAP and 68\% C.I. region for $T_{\r{excess},z}$ with black solid line and blue shaded region, respectively.
It can be seen that,
at 1 sigma (68\%) credible interval
,
the upper bound on $T_{\r{excess},z}$ exhibits roughly 10\% deviation relative to the MAP value.
Therefore in our ARCADE2 likelihood,
we set the uncertainty level $\sigma_\r{ARCADE2}$ to 10\% of the MAP $T_{\r{excess},z}$.

\section{Parameter posteriors}
\label{Appdx_CornerPlot}

In \Fig{Fig_Corner_Plot} we present the posteriors for astrophysical parameters and derived $\Tbb$ at $z=17$ for 3 representative inference models:
the highest evidence model (7 FG terms, with calibration residuals),
the lowest evidence model (4 FG terms, no calibration residuals),
and the {\tt Narrow} $f_\r{esc,7}$ pseudo-likelihood model from Appendix \ref{Sec_Pseudo_Likelihood}.

In the lowest evidence and pseudo-likelihood models, both of which have non-standard $\Tbb$ depths, the relevant parameters are tightly constrained.
Specifically,
$f_\r{R,III}$ is constrained to $f_\r{R,III} \gtrsim 10^{2.7}$,
which also marks the minimum radio efficiency required for Pop III radio excess background to cause noticeable deviation of $\Tbb$ from the standard regime.
Both models adjust the cosmic signal timing to match that of \citetalias{Bowman:2018yin} either through ionization ($f_\r{esc,7}$) or heating ($\LXIII$).  
In the lowest evidence model,
timing and sharp rise of $\Tbb$ are adjusted via increased ionization,
and the $f_\r{esc,7}$ posterior peaks close to our prior upper bound of $10^{-1}$,
resulting in an optical depth of $\tau_\r{rei} = 0.165$.
Whereas in the pseudo-likelihood for which a high $f_\r{esc,7}$ is not viable due to the prior,
a higher $\LXIII$ is required in order to match \citetalias{Bowman:2018yin} timing.

In comparison,
for our highest evidence model,
EDGES data structure is mostly explained by FG and calibration systematics rather than the cosmic signal.
 Thus the constraints on astrophysical parameters are very weak.
Since EDGES data intrinsically disfavors a strong non-standard $\Tbb$ signal, we only recover upper limits on $f_\r{R,III}$ and $f_{\star, 7}$.

\label{lastpage}
\end{document}